\documentclass[a4paper,authoryear,11pt]{article}

\usepackage{amsmath,amsfonts,amssymb,latexsym,wasysym,mathrsfs}
\usepackage{amsthm}
\usepackage{verbatim}
\usepackage[english]{babel}
\usepackage{a4wide}
\usepackage{pstricks,graphicx,subfigure}

\theoremstyle{plain}
\newtheorem{theorem}{Theorem}[section]
\newtheorem{proposition}{Proposition}[section]
\newtheorem{lemma}{Lemma}[section]
\newtheorem{corollary}{Corollary}[section]

\theoremstyle{definition}
\newtheorem{definition}{Definition}[section]
\newtheorem{fact}{Fact}[section]

\def \R {{\mathbb R}}

\def \ra {\rightarrow }

\def \E{{\mathbb{E}}}

\def \e {\epsilon}

\def \d {\delta}

\def \b {\beta}

\def \a {\alpha}

\def \ra {\rightarrow}

\newcommand{\be}[1]{\begin{equation}\label{#1}}
\newcommand{\ee}{\end{equation}}
\newcommand{\bl}[1]{\begin{lemma}\label{#1}}
\newcommand{\br}[1]{\begin{remark}\label{#1}}
\newcommand{\brs}[1]{\begin{remarks}\label{#1}}
\newcommand{\bt}[1]{\begin{theorem}\label{#1}}
\newcommand{\bd}[1]{\begin{definition}\label{#1}}
\newcommand{\bp}[1]{\begin{proposition}\label{#1}}
\newcommand{\bc}[1]{\begin{corollary}\label{#1}}
\newcommand{\bfact}[1]{\begin{fact}\label{#1}.}
\newcommand{\bex}[1]{\begin{example}\label{#1}.}
\newcommand{\ec}{\end{corollary}}
\newcommand{\efact}{\end{fact}}
\newcommand{\eex}{\end{example}}
\newcommand{\el}{\end{lemma}}
\newcommand{\er}{\end{remark}}
\newcommand{\ers}{\end{remarks}}
\newcommand{\et}{\end{theorem}}
\newcommand{\ed}{\end{definition}}
\newcommand{\ep}{\end{proposition}}
\newcommand{\epr}{\end{proof}}
\newcommand{\bpr}{\begin{proof}}
\newcommand{\bcl}[1]{\begin{claim}\label{#1}}
\newcommand{\ecl}{\end{claim}}

\newcommand{\bi}{\begin{itemize}}
\newcommand{\ei}{\end{itemize}}
\newcommand{\ben}{\begin{enumerate}}
\newcommand{\een}{\end{enumerate}}

\newcommand{\sign}{\text{sign}}

\usepackage{tikz}
\newcommand*\circled[1]{\tikz[baseline=(char.base)]{
            \node[shape=circle,draw,inner sep=.5pt] (char) {#1};}}

\begin{document}

\title{Polarization and coherence in mean field games driven by private and social utility}

\author{Paolo Dai Pra\footnote{University of Verona, Department of Computer Science, 15, Strada Le Grazie, I - 37134 Verona, Italy; contact:
 {paolo.daipra@univr.it}} \and Elena Sartori\footnote{University of Padova, Department of Mathematics ``Tullio Levi-Civita'', 
63, Via Trieste, I - 35121, Padova, Italy; contact: {esartori@math.unipd.it}} \and  Marco Tolotti\footnote{Ca' Foscari University of Venice, Department of Management, Cannaregio 873, I - 31121, Venice, Italy; contact: {tolotti@unive.it}}}

\date{}

\maketitle

\begin{abstract} 
We study  a mean field game in continuous time over a finite horizon, T, where the state of each agent is binary and where players base their strategic decisions on two, possibly competing,  factors: the willingness to align with the majority (conformism) and the aspiration of sticking with the own type (stubbornness). We also consider a quadratic cost related to the rate at which a change in the state happens: changing opinion may be a costly operation. Depending on the parameters of the model, the game may have more than one Nash equilibrium, even though the corresponding  N-player game does not.  Moreover, it exhibits a very rich phase diagram, where polarized/unpolarized, coherent/incoherent equilibria may coexist, except for T small, where the equilibrium is always unique. We fully describe such phase diagram in closed form and provide a detailed numerical analysis of the $N$-player counterpart of the mean field game. In this finite dimensional setting, the equilibrium selected by the population of players is always coherent (favoring the subpopulation whose type is aligned with the initial condition), but it does not necessarily minimize the cost functional. Rather, it seems that, among the coherent ones, the equilibrium prevailing is the one that most benefits the {\it underdog} subpopulation forced to change opinion.
\bigskip

\noindent \textbf{Keywords}: Finite Population Dynamics; Mean field games; Multiple Nash Equilibria; Phase transition; Social Interaction

\bigskip
\noindent \textbf{JEL Classification}: C61; C73; D91

\medskip
\noindent \textbf{MSC Classification}: 91A16; 91B14



\end{abstract}

\newpage

\section{Introduction}

In this paper we analyze a simple continuous-time dynamic multi-agent model, and study the limit as the number of agents goes to infinity. 
We consider a group of $N$ interacting agents ({\em players}), who are allowed to control their {\em binary} state choosing the probability rate of ``flipping'' them. This rate is a feedback control, that may depend on the state of all players and (measurably) on time. Each player aims at minimizing an individual cost, which is comprised by a {\em running cost} and a {\em final reward}. We consider a standard quadratic running cost. At some final time $T>0$, each player gets a reward given as the sum of two different terms: 
\bi
\item the first mimics a \emph{social driver} and favors imitation: the player gets a higher reward if she conforms with the majority, and if the majority becomes polarized (close to consensus); the majority is over the whole population, making the interaction between players of {\em mean field} type;
\item the second models the private (individual) desire to align the state with the sign of a static and predetermined random variable denoting her personal type. 
\ei
These two terms are possibly competing and represent a classical social dilemma: the former mimics \emph{conformism}, i.e., the adherence to social norms and is often referred to as social utility; the latter models \emph{stubbornness}, namely, the aspiration of the agent to stay as close as possible to the prescription of personal traits, hence mimicking a private (or individual) utility. As notion of optimality we adopt that of {\em Nash equilibrium}, and our aim is to understand the system's behavior in the limit as $N \ra +\infty$. This falls into the realm of {\em mean field games}, introduced by J.-M.~Lasry and P.-L.~Lions and, independently, by M.~Huang, R.P.~Malham{\'e} and P.E.~Caines (cf.\ \cite{lasrylions07}, \cite{huangetal06}), as limit models for symmetric  many player dynamic games as the number of players tends to infinity; see, for instance, the lecture notes \cite{cardaliaguet13} and the two-volume work \cite{carmonadelarue}. 

The variable representing the type is introduced in the model as a random field and is treated as an {observable} static component of player' state. Therefore, this term introduces random disorder and, to the best of our knowledge, it is one of the first attempts to do it in mean field games.

In literature, this dilemma has been analyzed from different perspectives: \cite{BeD} is usually considered as a pioneering study on the trade-off between private and social drivers in (static) binary choice models. In \cite{BlumeDurlauf}, a generalization to a dynamic continuous-time setting is proposed, which is not a mean field game, as agents play static games at random times. \cite{DST} proposes a model of consensus formation similar to our, where only the social component is present and individual preferences are not considered. \cite{BPT} is one of the rare examples studying the interplay between stubbornness and imitation drivers in the realm of mean field games. However, their mathematical setting is rather different from ours: in that paper state variables are real with Gaussian initial distribution; moreover, their linear quadratic optimization problem is solved by affine controls which preserve Gaussianity; as a consequence, their optimal control is always unique. Models close to the one proposed here, but without individual preferences, have also been introduced as examples of non-uniqueness of equilibria in mean field games (see, e.g., \cite{BZ20,BF19,CDFP19,NST20,DT20,haiek2019}).

Similarly to what is contained in the last cited references, the model that we are proposing here, has the following remarkable feature: for each $N$, there is a unique Nash equilibrium for the $N$-player game; however, the corresponding mean field game may have multiple equilibria. This is reminiscent of a common paradigm in Statistical Physics: finite volume Gibbs states are uniquely defined, but thermodynamic limit may be non unique, indicating a {\em phase transition}. The analogy with models in Statistical Physics can be carried on further: the model that we propose corresponds to the mean field Ising model (or Curie-Weiss one), when there are no private signals/types, and to the random field Curie-Weiss model, when disorder is introduced. The time horizon $T$ plays a role similar to the inverse temperature in the models cited above: the higher $T$, the smaller the contribution of the running cost. The $N$-player game as well as its mean field limit are presented in Section \ref{sec:model}. In remarkable analogy with the Curie-Weiss model, we show that the mean field game has a unique equilibrium for small $T$, whereas several equilibria emerge as $T$ increases.\\
\indent
A detailed study of the different equilibria of the mean field limit is proposed in Section \ref{sec:equilibria}. In particular, we will see that a number of different types of equilibria can be identified: polarized/unpolarized (related to the size of the majority), coherent/incoherent (alignment of the final population state with the initial state). It is, therefore, natural to ask which one of these equilibria is selected by taking the limit of the unique equilibrium for the N-player game. In \cite{CDFP19}, in absence of individual preferences and with a much simpler phase diagram, this question was rigorously answered, while a rigorous analysis is presently out of reach here. We, therefore, run numerical simulations to capture this selection. We see that  there is, indeed, a unique equilibrium emerging from the N-player approximation: it is always coherent, but it could be polarized or not, depending on some parameters of the model. Notably, the prevailing equilibrium is not necessarily the one that minimizes the aggregate cost suffered by the population of interacting agents. Some remarks about the rationale behind the selection of the equilibrium in the case of a finite population are collected in Section \ref{sec:N_economic}. Section \ref{sec:conclusion} contains some concluding remarks. The Appendix contains all technical proofs of the results stated in Section \ref{sec:equilibria}.

\section{A continuous-time binary strategic game}\label{sec:model}

\subsection{The $N$-player game}

We consider $N$ players whose binary state vector is denoted by ${\bf x} := (x_1, \ldots,x_N) \in \{-1,1\}^N$. To each player is also assigned a variable $y_i \in \{-\e,\e\}$, where $\e>0$ is a given constant, and we set ${\bf y} := (y_1,\ldots,y_N)$; the components of ${\bf y}$ will be referred to as {\em local fields}. The vector state ${\bf x} = {\bf x}(t)$ evolves in continuous time, while ${\bf y}$ is static. Each player is allowed to control her state with a feedback control $u_i(t,{\bf x}, {\bf y})$ which may depend on time, and on the values of ${\bf x}$ and ${\bf y}$. We assume each $u_i$, as function of $t$, to be nonnegative, measurable and locally integrable. Thus, for a given control ${\bf u} = (u_1, \ldots, u_N)$, the state of the system, ${\bf x}(t)$, evolves as a Markov process, whose law is uniquely determined as the solution of the martingale problem for the time-dependent generator
\[
\mathcal{L}_t f({\bf x}) := \sum_{i=1}^N u_i(t,{\bf x}, {\bf y}) \left[f({\bf x}^i) - f({\bf x}) \right] =:  \sum_{i=1}^N u_i(t,{\bf x}, {\bf y}) \nabla^i f({\bf x}),
\]
where ${\bf x}^i$ is the vector state obtained from ${\bf x}$ by replacing the component $x_i$ with $-x_i$. In order to fully define the dynamics, we prescribe the joint distribution of the initial states ${\bf x}(0)$ and of the local fields ${\bf y}$. For simplicity we assume all these variables are independent: all $x_i(0)$ have mean $m_0 \in [-1,1]$, whereas all $y_i$ have mean $0$.
Each player aims at minimizing  her own cost, depending on the controls of all players, which is given by
\be{costi}
J_i({\bf u}) := \E \left[ \frac12\int_0^T u_i^2(t, {\bf x}(t), {\bf y})dt - x_i(T) (m_N(T) + y_i) \right],
\ee
where $m_N(t) := \frac{1}{N} \sum_{i=1}^N x_i(t)$, is the mean state of the population.
Here, $T>0$ is the time horizon of the game. Besides the standard quadratic running cost in the control, two other terms contribute to the cost:
\bi
\item
the term $-x_i(T) m_N(T)$ favors \emph{polarization}: each agent profits from being aligned with the majority at the final time $T$;
\item
the term  $-x_i(T) \, y_i$ incentivizes each agent to align with her own local field. As ${\bf y}$ is uniformly distributed on $\{-\e,\e\}^N$, this term inhibits alignment of behaviors, hence polarization.
\ei
From a technical viewpoint, we note that, by rescaling time, one could normalize to $1$ the time horizon and multiplying by $T$ the reward. Thus, the time horizon $T$ may be seen as tuning the relevance of the final reward as compared to the ``natural inertia'' expressed by the running cost. 

Given a control vector ${\bf u}$ and a measurable and locally integrable function
\[
\b: [0,T] \times \{-1,1\}^N \times \{-\e,\e\}^N \ra [0,+\infty),
\]
we define the control vector $\left[{\bf u}^i,\b\right]$ by
\[
\left[{\bf u}^i,\b\right]_j = \left\{ \begin{array}{ll} u_j & \mbox{for } j \neq i \\ \b & \mbox{for } j=i \end{array} \right..
\]

\begin{definition}\label{def:Nash}
A control vector ${\bf u}$ is a {\em Nash equilibrium} if for each $\b$ as above,
\[
J_i({\bf u}) \leq J_i(\left[{\bf u}^i,\b\right]),\quad \forall i =1,\ldots,N.
\]
\end{definition}
\noindent
Nash equilibria may be obtained via the Hamilton-Jacobi-Bellman equation (see for details \cite{DJVS})
\be{hjbN}
\begin{cases}
\frac{\partial {v_i}}{\partial t}(t,{\bf x}, {\bf y}) + \sum_{j=1}^N a_*(\nabla^j v_j(t,{\bf x}, {\bf y}) ) \nabla^j v_i(t,{\bf x}, {\bf y}) + \frac12 a_*^2( \nabla^i v_i(t,{\bf x}, {\bf y})) = 0 \\
v_i(T, {\bf x}, {\bf y}) = -x_i(m_N+y_i),
\end{cases}
\ee
where $\nabla^j v_j (t,{\bf x}, {\bf y})=v_j (t,{\bf x}^j, {\bf y})-v_j (t,{\bf x}, {\bf y})$, $m_N=\frac{1}{N}\sum_{i=1}^N x_i$, and 
\[
a_*(p) := \arg\min_{a \geq 0} \left[ ap - \frac12 a^2 \right] = p^-
\]
with $p^-$ denoting the negative part of $p \in \R$. Note that \eqref{hjbN} is a system of $N\times 2^N\times 2^N$ ordinary differential equations with locally Lipschitz vector field and global solutions. Therefore, it admits a unique solution ${\bf v} := (v_1,\ldots,v_N)$; moreover, there exists a unique Nash equilibrium ${\bf u}$ given by
\[
u_i(t,{\bf x}, {\bf y}) := a_*( \nabla^i v_i(t,{\bf x}, {\bf y})),\quad i=1,\ldots,N.
\]

\subsection{The mean field game}

The mean field game is the formal limit of the above $N$-player game, as seen from a representative player. Denote respectively by $x \in \{-1,1\}$ and $y\in \{-\e,\e\}$ the state and the local field of a representative player. Given a (deterministic)  $m\in [-1,1]$, the player aims at minimizing the cost 
\be{costmfg}
J(u) := \E\left[\frac{1}{2} \int_0^T u^2(t,x(t), y)dt - x(T)(m + y) \right]
\ee
under the Markovian dynamics with infinitesimal generator
\be{genmfg}
L^u_t f(x,y) := u(t,x,y) [f(t,-x,y) - f(t,x,y)].
\ee
As above, admissible controls are measurable and locally integrable functions. Consistently with the $N$-player game, the initial state $x(0)$ and the local field $y$ are independent, with means $m_0$ and $0$, respectively. As we will see below, the convexity of the running cost implies uniqueness of the optimal control $u_*^m$, which actually depends on the choice of $m$. Denoting by $x^m_*(t)$ the evolution of the state for the optimal control, the solution of the mean field game is completed by finding the solutions of the Consistency Equation
\be{consistency}
m = \E\left(x^m_*(T)\right).
\ee
Therefore, we handle the mean field game first by solving, for $m$ fixed, the optimal control problem \eqref{costmfg}-\eqref{genmfg} via Dynamic Programming. This leads to the Hamilton-Jacobi-Bellman equation
\be{hjbmfg}
\begin{cases}
\frac{\partial V(t,x,y)}{\partial t} + \min_{u \geq 0} \left[ u \nabla V(t,x,y) + \frac12 u^2 \right] = 0 \\
V(T,x,y) = -x(m+y)
\end{cases}
\ee
with $\nabla V(t,x,y) := V(t,-x,y) - V(t,x,y)$. 
The optimal feedback control is given by 
\[
u_*^m(t,x,y) = \left(\nabla V(t,x,y)\right)^-.
\]
Then we solve \eqref{consistency} using \eqref{genmfg} to obtain the evolution of $m(t,y) :=  \E_y\left(x^m_*(t)\right)$, where $\E_y$ denotes expectation conditioned to the local field $y \in \{-\e,\e\}$. It follows that $ \E\left(x^m_*(t)\right) = \frac12 \left[ m(t,\e) + m(t,-\e) \right]$.
By \eqref{genmfg}, we obtain the Kolmogorov forward equation
\be{kolm}
\frac{d}{dt}m(t,y) =  \E_y\left[-2 u_*(t,x^m_*(t), y) x^m_*(t)\right].
\ee
Now we proceed with the explicit solutions of the steps just described. Next, we discuss the solutions of the Consistency Equation \eqref{consistency} in terms of the three parameters of the model: $T$, $\e$ and $m_0$.

\subsection{Solving the Hamilton-Jacobi-Bellman equation}

Our aim here is to determine the value function $V(t,x,y)$ which solves \eqref{hjbmfg}. Note that this value function also depends on $m$. It is convenient to set
$
z(t,y) := V(t,-1,y) - V(t,1,y).
$
Note also that 
$
\nabla V(t,x,y) = x z(t,y).
$
Using \eqref{hjbmfg}, we can subtract the two equations for $V(t,-1,y)$ and $V(t,1,y)$, obtaining a closed equation for $z(t,y)$:
\be{hjbz}
\begin{cases}
\frac{d}{dt} z(t,y) = \frac12 z(t,y) |z(t,y)| \\
z(T,y) = 2(m+y).
\end{cases}
\ee
It is a key fact that the equations for $z(t,\e)$ and $z(t,-\e)$ are decoupled, so they can be solved separately by separation of variables. Indeed, observing that, by uniqueness, the sign of $z(t,y)$ is constant in $t\in [0,T]$, we can rewrite \eqref{hjbz} as
\[
 \frac{d}{dt} \left( \frac{1}{z(t,y)} \right)= \frac12 \sign(m+y)
 \]
 that, integrated over $[t,T]$, yields
 \be{zt}
 z(t,y) = \frac{2(m+y)}{|m+y|(T-t) + 1}.
 \ee
 At this point we can also compute the value function $V(t,x,y)$. Plugging the optimal control in \eqref{hjbmfg} we get
 \[
 \frac{\partial}{\partial t} V(t,1,y) = \frac12 \left(z^-(t,y)\right)^2 = \left\{ \begin{array}{ll} 0 & \quad\mbox{ if } m+y \geq 0 \\ 
 \left( \frac{2(m+y)}{|m+y|(T-t) + 1} \right)^2 & \quad\mbox{ if } m+y<0. \end{array} \right.
 \]
 \\
Integrating this last identity from $t$ to $T$ we get $V(t,1,y)$. \\
Having $V(t,1,y)$ and \mbox{$z(t,y) = V(t,-1,y) - V(t,1,y)$} we can also obtain $V(t,-1,y)$. The final result is 
\begin{equation}\label{eq:v}
 V(t,x,y) = \left\{ \begin{array}{ll} -|m+y| &\quad \mbox{ if } \sign(m+y) \in \{0,x\} \\ -|m+y| + \frac{2|m+y|}{|m+y|(T-t)+1} & \quad \mbox{ if } \sign(m+y) = -x. \end{array} \right.
 \end{equation}
 
 \subsection{Solving the Kolmogorov forward equation}
 
 We begin by observing that 
 \[
 u_*(t,x,y) = \left(\nabla V(t,x,y)\right)^- =  \left(x z(t,y)\right)^- = \frac{1+x}{2} z^-(t,y)  + \frac{1-x}{2} z^+(t,x) = 
  \frac12 |z(t,y)| - \frac{x}{2} z(t,y).
 \]
 Plugging this into \eqref{kolm} we obtain  
 \[
 \begin{cases}
 \frac{d}{dt}m(t,y) = - m(t,y) |z(t,y)| + z(t,y) \\ m(0,y) = m_0,
 \end{cases}
 \]
 where we used the fact that $x^2 = 1$.
 Recalling that the sign of $z(t,y)$ is constant and equals $\rho:= \sign(m+y)$ we can rewrite this last equation as
 \[
 - \rho \frac{d}{dt}\log(1-\rho m(t,y))  = z(t,y) =  \frac{2(m+y)}{\rho(m+y)(T-t) + 1},
 \]
 which, integrated from $0$ to $t$, yields
 \be{mty}
 m(t,y) = \rho \left[ 1-(1-\rho m_0) \left( \frac{|m+y|(T-t)+1}{|m+y|T+1} \right)^2 \right].
 \ee
Hence,
 \be{mean}
 \E\left(x_*^m(t) \right) = \frac12 \left( m(t,\e) + m(t,-\e)\right).
 \ee

\section{Equilibria and phase diagram}\label{sec:equilibria}

This section is entirely devoted to the analysis of the solutions to the Consistency Equation \eqref{consistency}. As said, $m$ corresponds to a solution of the MFG if and only if it solves such equation. 
Relying on \eqref{mty} and \eqref{mean},   the Consistency Equation  can be rewritten as  
\be{eq}
\frac12 \sign(m+\e)\left[1-\frac{(1-\sign(m+\e)m_0)}{(|m+\e|T+1)^2}\right]  + \frac12 \sign(m-\e)\left[1-\frac{(1-\sign(m-\e)m_0)}{(|m-\e|T+1)^2}\right] = m.
\ee
Solutions of \eqref{eq} will be called {\em equilibria}.  We are now going to  identify all such equilibria. Moreover, depending on the values of the parameters of the model, we classify them emphasizing the presence or the absence of two different features:
\bi
\item[-] \emph{polarization}, which expresses the fact that agent alignment outscores individual preference,
\item[-] \emph{coherence}, which indicates the fact that the majority of the agents aligns with the  sign of the initial condition $m_0$.
\ei

We begin by pointing out a symmetry property:
if we denote by $F(m,\e,T,m_0)$ the l.h.s. of \eqref{eq}, then
\be{simmF}
F(-m,\e,T,-m_0) = F(m,\e,T,m_0).
\ee
Therefore, without loss of generality, in the remainder of this article we study \eqref{eq} in the case $m_0 \geq 0$.  We can thus specify precisely four classes of equilibria:
\bi
\item equilibria $m \in (\e,1]$ will be called {\em polarized coherent};
\item equilibria $m \in [-1,-\e]$ will be called {\em polarized incoherent};
\item equilibria $m \in (0,\e]$ will be called {\em unpolarized coherent};
\item equilibria $m \in [-\e,0]$ will be called {\em unpolarized incoherent}.
\ei

\begin{figure}[ht]
{\includegraphics[width=.5\columnwidth]{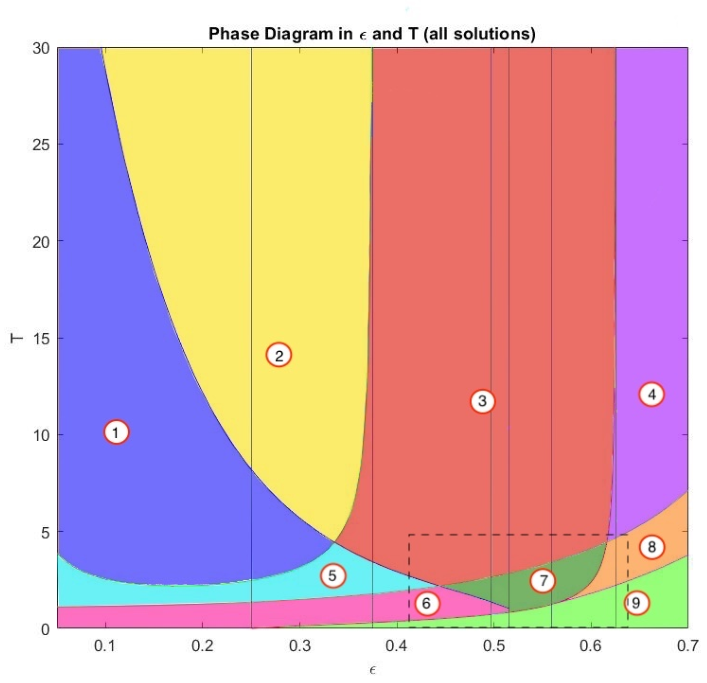}}
{\includegraphics[width=.5\columnwidth]{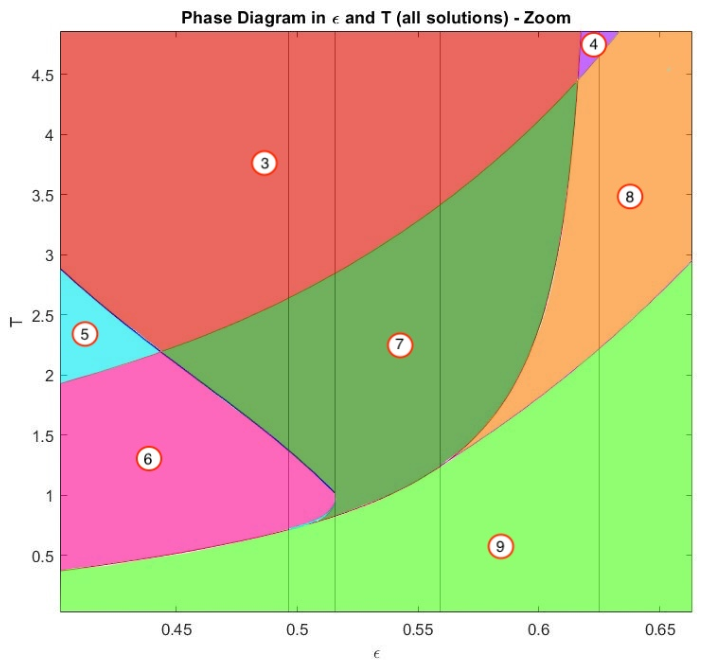}}
\caption{Full phase diagram for $m_0 = 0.25$. Different regions identify different properties of the solutions to \eqref{eq}, both in terms of numerosity and classification.
}   
\label{fig1}
\end{figure}

Before stating the formal results describing in detail all the solutions to \eqref{eq}, we provide a visual example, where all the possible situations are depicted. To this aim, in Figure \ref{fig1} we plot the full phase diagram in the parameters $\e,T$, having fixed $m_0 = 0.25$. The right picture is a zoom of the region within the dashed lines. We can identify nine regions, each of them corresponding to a specific typology of solutions to \eqref{eq}. For example, \emph{region 1}, characterized by an intermediate value of $T$ and $\e$ small, shows the presence of three equilibria, one polarized/coherent, one polarized/incoherent, and one unpolarized/incoherent.
\begin{table}[ht]
   \label{tab:eq}
\vspace{.1 in}
  \begin{center}
  \begin{tabular}[]{c|c|cccc}
 Region & Number of &  Polarized & Polarized & Unpolarized & Unpolarized\\
& equilibria & Coherent  &  Incoherent  &  Coherent &  Incoherent\\
\hline
   \circled{1}  & 3 & 1 & 1 & 0 & 1\\
   \circled{2}  & 5 & 1 & 1 & 2 & 1\\
   \circled{3}  & 5 & 1 & 2 & 2 & 0\\
   \circled{4}  & 5 & 2 & 2 & 1 & 0\\
   \circled{5}  & 3 & 1 & 2 & 0 & 0\\
   \circled{6}  & 1 & 1 & 0 & 0 & 0\\
   \circled{7}  & 3 & 1 & 0 & 2 & 0\\
   \circled{8}  & 3 & 2 & 0 & 1 & 0\\
   \circled{9}  & 1 & 0 & 0 & 1 & 0\\
	  \end{tabular}\end{center}
	  \caption{Total number of equilibria for each region of the phase diagram as in Figure \ref{fig1},  and number of equilibria disentangled for their typology (coherent/incoherent; polarized/unpolarized).}
 \end{table}

In Table \ref{tab:eq}, we summarize the results in terms of number of equilibria and their typology for all the regions on the phase diagram as depicted in Figure \ref{fig1}.  We note that in regions 6 and 9, for $T$ small, there is a unique equilibrium which is always coherent, and it is polarized in 6, for $\e$ small, whereas it is unpolarized in 9, for $\e$ large. On the opposite, in zones 2, 3 and 4, for $T$ large, there are five equilibria, and three of them are always coherent. Finally, in zones 1, 5, 7 and 8 (for $T$ intermediate), there are three equilibria. In this situation, we see two different zones: in regions 1 and 5 (for $\e$ small), only one equilibrium is coherent; in regions 7 and 8 (for $\e$ large), they are all coherent.

Note that the way the number of equilibria depends of the parameters $\e$ and $T$ is far from obvious. For instance, fixing $\e \simeq 0.5$, the number of equilibria is not monotonic in $T$. Similarly, fixing $T \simeq 3$, there is no monotonicity in $\e$.

In the remainder of this section, we state the results describing the phase diagram, specifying at what times the phase transitions occur. This will also serve to specify the algebraic form of all the curves separating the regions depicted in Figure \ref{fig1}.  To ease readability,  we organize them in four propositions, one for each type of equilibrium of the MFG, i.e., one for each class identified by the possible polarization or coherence of the equilibria, which are listed in Table \ref{tab:eq}. As already mentioned, we restrict to the case $m_0\geq 0$. 


\begin{proposition}[Polarized coherent equilibria: $m > \e$]\label{prop:equilibria} \phantom{...}\\
\vspace{-.5cm}
\bi
\item[(i)] 
Suppose $\e \leq m_0$. Then, $\forall T>0$, there is a unique equilibrium $m = M(T,\e,m_0)$ in $(\e,1]$, [regions 6, 5, 1, 2 in Fig. \ref{fig1}]. Moreover,
$
\lim_{T \ra +\infty} M(T,\e,m_0) = 1.
$
\item[(ii)]
Suppose $\e \geq \frac{1+m_0}{2}$. Then there exists a unique $T^{(1)}_c= T^{(1)}_c(\e,m_0)>0$ such that the graph of the curve in the plane $(z,m)$ of equation $z = F(m,\e, T^{(1)}_c, m_0)$ is tangent to the line of equation $z=0$. Moreover,
\bi
\item
for $T < T^{(1)}_c(\e,m_0)$, there is no equilibrium in $(\e,1]$, [region 9];
\item
for $T =T^{(1)}_c(\e,m_0)$, there is a unique equilibrium in $(\e,1]$, [separatrix of regions 9 and 8];
\item
for $T >T^{(1)}_c(\e,m_0)$, there are two equilibria in $(\e,1]$, [regions 8, 4].
\ei
\item[(iii)]
Suppose $m_0 < \e < \frac{1+m_0}{2}$. Define
\be{Tstar}
T^*(\e, m_0) := - \frac{1}{2\e} + \frac{1}{2\e} \sqrt{\frac{1-m_0}{1+m_0-2\e}}>0
\ee
Then there exists a unique $\e_*^{(1)} \in \left(m_0,\frac{1+m_0}{2}\right)$ such that 
$
\frac{\partial}{\partial m} F(\e,\e,T^*(\e, m_0),m_0) = 0.
$
\\
Moreover, if $m_0 < \e \leq \e_*^{(1)}\!,$
\bi
\item
for $T \leq T^*(\e, m_0)$, there is no equilibrium in $(\e,1]$, [region 9];
\item
for $T > T^*(\e, m_0)$, there is a unique equilibrium in $(\e,1]$, [regions 6, 7, 5, 3, 1, 2].
\ei

If $\e_*^{(1)} < \e < \frac{1+m_0}{2}$, there exists a unique $T^{(1)}_c(\e,m_0) >0$ defined as for the case \mbox{$\e \geq \frac{1+m_0}{2}$. }Furthermore, $T^{(1)}_c(\e,m_0) <  T^*(\e, m_0)$,  for each $m_0$ the map $\e \mapsto T_c^{(1)}(\e,m_0)$ is continuous, $T_c^{(1)}(\e,m_0) \ra T^*(\e_*^{(1)}, m_0)$ as $\e \downarrow \e_*^{(1)}$, $T_c^{(1)}$ defined here for $\e\in(\e_*^{(1)},\frac{1+m_0}{2})$ connects continuously at $\e= \frac{1+m_0}{2}$ with $T_c^{(1)}$ defined for $\e \geq \frac{1+m_0}{2}$, and
\bi
\item
for $T<T_c^{(1)}(\e,m_0)$, there is no equilibrium in $(\e,1]$, [region 9];
\item
for $T=T_c^{(1)}(\e,m_0)$, there is a unique equilibrium in $(\e,1]$, [separatrix of 9 and 8];
\item
for $T_c^{(1)}(\e,m_0)<T<T^*(\e, m_0)$, there are two equilibria in $(\e,1]$, [regions 8, 4];
\item
for $T \geq T^*(\e, m_0)$, there is a unique equilibrium in $(\e,1]$, [regions 7, 3].
\ei

\ei

\end{proposition}


\begin{proposition}[Polarized incoherent equilibria: $m < -\e$]\label{prop:equilibria2}\phantom{...}\\
\vspace{-.5cm}
\bi
\item[(i)]
Suppose $\e \geq \frac{1-m_0}{2}$. Then, 
\bi
\item
for $T < T^{(1)}_c(\e,-m_0)$, there is no equilibrium  in $[-1,-\e)$, [regions 9, 6, 7, 8];
\item
for $T =T^{(1)}_c(\e,-m_0)$, there is a unique equilibrium  in $[-1,-\e)$, [separatrix of  6, 7, 8 from 5, 3, 4];
\item
for $T >T^{(1)}_c(\e,-m_0))$, there are two equilibria  in $[-1,-\e)$, [regions 5, 3, 4].
\ei
\item[(ii)]
Suppose $0< \e < \frac{1-m_0}{2}$. Then $T^{(1)}_c(\e,-m_0) < T^{(1)}_*(\e,-m_0)$ , and,
\bi
\item
for $T<T^{(1)}_c(\e,-m_0)$, there is no equilibrium in $[-1,-\e)$, [region 6];
\item
for $T=T^{(1)}_c(\e,-m_0)$, there is a unique equilibrium in $[-1,-\e)$, [separatrix of 6 and 5];
\item
for $T^{(1)}_c(\e,-m_0)<T<T^{(1)}_*(\e,-m_0)$, there are two equilibria in $[-1,-\e)$, \mbox{[region 5];}
\item
for $T \geq T^*(\e,-m_0)$, there is a unique equilibrium  in $[-1,-\e)$, [regions 1, 2].
\ei
\ei

\end{proposition}

\begin{proposition}[Unpolarized coherent equilibria: $0\leq m\leq \e$]\label{prop:equilibria3}\phantom{...}\\
\vspace{-.5cm}
\bi
\item[(i)]
Suppose $\e \leq m_0$. Then there exists a unique $T^{(2)}_c= T^{(2)}_c(\e,m_0)>0$ such that the graph of the curve in the plane $(z,m)$ of equation $z = F(m,\e, T^{(2)}_c, m_0)$ is tangent to the line of equation $z=0$.  Moreover,
\bi
\item
for $T < T^{(2)}_c(\e,m_0)$, there is no equilibrium in $(0,\e)$, [regions 6, 5, 1];
\item
for $T =T^{(2)}_c(\e,m_0)$, there is a unique equilibrium in $(0,\e)$, [separatrix of 1 and 2];
\item
for $T >T^{(2)}_c(\e,m_0)$, there are two equilibria in $(0,\e)$, [region 2].
\ei
\item[(ii)]
Suppose $\e \geq \frac{1+m_0}{2}$. Then, $\forall T>0$, there is a unique equilibrium $m = M(T,\e,m_0)$ in $(0,\e)$, [regions 9, 8, 4]. Moreover, 
$
\lim_{T \ra +\infty} M(T,\e,m_0) = 0.
$
\item[(iii)]
Suppose $m_0 < \e <  \frac{1+m_0}{2}$ and let $T^*(\e,m_0)$ given by \eqref{Tstar}. \\ For $T \leq T^*(\e,m_0)$ , there is a unique solution in $(0,\e]$, [regions 9, 8, 4]; the solution is $\e$ if and only if \mbox{$T = T^*(\e,m_0)$.} Consider now the function
\begin{multline*}
V(s,\e,m_0):= \frac{\sqrt{s}}{8} \left\{ 16 + \sqrt{s} \left[ 16 + 3 m_0\left(\frac{2m_0}{\e s} \right)^{1/3} \right] \right\} \\ + \frac{\e}{2} (1+\sqrt{s})^2 \left\{ -2 - 4 \sqrt{s} + s \left[ -2+3 \left(\frac{2m_0}{\e s} \right)^{1/3} \right] \right\}.
\end{multline*}
Then,
\bi \item
the equation for the unknown $\e,$ 
$
V\left(\frac{m_0}{4 \e},\e,m_0\right) = 0
$ 
has a unique solution\\ \mbox{$\e_*^{(2)}(m_0) \in \left(m_0, \frac{1+m_0}{2} \right)$} (unless $m_0 = 0$ for which $\e_*^{(2)}(m_0) = 0$);
\item
there is a unique $\e_*^{(3)}(m_0) \in \left(\e_*^{(2)}(m_0), \frac{1+m_0}{2} \right)$ such that the curve in the plane $(s,z)$ of equation $z = V(s,\e_*^{(3)}(m_0),m_0)$ is tangent to the line of equation $z=0$.
\ei
Moreover,
\bi
\item[(a)] 
if $m_0 < \e \leq \e_*^{(2)}(m_0)$, the critical time $T^{(2)}_c= T^{(2)}_c(\e,m_0)>0$ defined in point (i) is well defined, and $T^*(\e,m_0) < T^{(2)}_c(\e,m_0)$;
\bi
\item[-]
for $T^*(\e,m_0) < T \leq T^{(2)}_c(\e,m_0)$, there is no equilibrium in $(0,\e]$, [regions 6, 5, 1];
\item[-]
for $T =T^{(2)}_c(\e,m_0)$, there is a unique equilibrium in $(0,\e)$, [separatrix of  1 from 2, 3];
\item[-]
for $T> T^{(2)}_c(\e,m_0)$, there are two equilibria in $(0,\e)$, [regions 3, 2];
\ei
\item[(b)]
if $\e_*^{(2)}(m_0) < \e < \e_*^{(3)}(m_0)$, $T^{(2)}_c$ is not well defined. This implies that there are two times \mbox{$T^{(2)}_c(\e,m_0) < T^{(3)}_c(\e,m_0)$} such that the graph of the curve in the plane $(z,m)$ of equation $z = F(m,\e, T^{(2)}_c, m_0)$ is tangent to the line of equation $z=0$ and
\bi
\item[-]
for $T^*(\e,m_0) < T \leq T^{(2)}_c(\e,m_0)$, there are two equilibria in $(0,\e]$, [region 7 in Fig. \ref{fig1}];
\item[-]
for $T = T^{(2)}_c(\e,m_0)$, there is a unique equilibrium in $(0,\e)$, [separatrix of 7 and 6];
\item[-]
for $T^{(2)}_c(\e,m_0)<T< T^{(3)}_c(\e,m_0)$, there is no equilibrium in $(0,\e)$, [region 6];
\item[-]
for $T = T^{(3)}_c(\e,m_0)$, there is a unique equilibrium in $(0,\e)$, [separatrix of  6 and 7];
\item[-]
for $T > T^{(3)}_c(\e,m_0)$, there are two equilibria in $(0,\e])$, [region 3];
\ei
\item[(c)]
if $ \e_*^{(3)}(m_0) \leq \e < \frac{1+m_0}{2}$, there are two equilibria in $(0,\e])$, $\forall T > T^*(\e,m_0)$, [regions 7, 3].
\ei
\ei
\end{proposition}


\begin{proposition}[Unpolarized incoherent equilibria: $-\e \leq m < 0$]\label{prop:equilibria4}\phantom{...}\\
\vspace{-.5cm}
\bi
\item[(i)]
Suppose $\e \geq \frac{1- m_0}{2}$. Then there is no equilibrium in $[-\e,0)$,  $\forall T>0$, [regions 9, 6, 7, 8, 5, 3, 4].

\item[(ii)]
Suppose $\e < \frac{1-m_0}{2}$ and let $T^*(\e,m_0)$ be given by  \eqref{Tstar}. Then,
\bi
\item
for $T \leq T^*(\e,-m_0)$,  there is no equilibrium in $(-\e,0)$ and $m=-\e$ is an equilibrium if and only if $T=T^*(\e,-m_0)$, [regions 9, 6, 5, 3];
\item
for $T>T^*(\e,-m_0)$, there is a unique equilibrium in $[-\e,0)$, [regions 1, 2].
\ei

\ei

\end{proposition}

\section{The $N$-player game: HJB and numerical results}\label{sec:Nfinite}

In this section, we provide a different glance to the problem and consider again the representative agent in a setting where she is best responding to a population of $N$ opponents (note that, in doing this, the population is thus formed by $N+1$ players). We now derive a new \emph{large dimensional} HJB equation used to run simulations of a finite population in order to identify the emerging (unique) equilibrium in the finite dimensional model. This approach is inspired by \cite{haiek2019}. 

All parameters and variables are as described in the previous sections.  Recall that each agent $j$ is characterized by a predetermined \emph{local field} $y_j\in\{-\e,\, \e\}$, where $\e\in[0,1]$ and by a time-varying state variable $x_j(t)\in \{ -1,\, 1\}$. 
We take agent $i$ playing the role of the representative agent. Concerning the remaining population of $N$ players, we  introduce two summary statistics as the number of ``ones'' in the two subpopulations with different local fields (i.e., different $\e$). To this aim, we define 
$$n^+_N = \sum_{j\ne i} \mathbb I_{\{x_j=1\}} \mathbb I_{\{y_j=\e\}};\quad  
n^-_N = \sum_{j\ne i} \mathbb I_{\{x_j=1\}} \mathbb I_{\{y_j=-\e\}}; \quad n^{\e}_N =  \sum_{j\ne i} \mathbb I_{\{y_j=\e\}}.$$
Note that $n^{\e}_N$ is a static variable, whereas $n^{+}_N$ and $n^-_N$ change in time, and take values, respectively, in $\{0,\,1,\ldots,n_N^{\e}\}$ and $\{0,\,1,\ldots,N-n_N^{\e}\}$. By taking advantage of the symmetries of the model we search equilibrium controls that, for the representative player $i$, are feedback depending on the state $x_i$, on the local field $y_i$ and on the aggregate variables $n^+_N, n^-_N$ and $n^{\e}_N$, and symmetrically for all other players. We denote by $\a(x_i,y_i,n^+,n^-,n^{\e},t)$ the feedback control strategy of player $i$, while each other player $j \neq i$ uses the feedback control  {\small
\[
\b(x_j, y_j, n^+\! -\mathbb I_{\{x_j=1\}} \mathbb I_{\{y_j=\e\}} + \mathbb I_{\{x_i=1\}} \mathbb I_{\{y_i=\e\}}, n^-\! -\mathbb I_{\{x_j=1\}} \mathbb I_{\{y_j=-\e\}} + \mathbb I_{\{x_i=1\}} \mathbb I_{\{y_i=-\e\}}, n^{\e}\! - \mathbb I_{\{y_j=\e\}} + \mathbb I_{\{y_i=\e\}}, t)
\]
}
where, for instance, we have used the fact that 
\[
n^+ -\mathbb I_{\{x_j=1\}} \mathbb I_{\{y_j=\e\}} + \mathbb I_{\{x_i=1\}} \mathbb I_{\{y_i=\e\}} = \sum_{k\ne j} \mathbb I_{\{x_k=1\}} \mathbb I_{\{y_k=\e\}}\,.
\]
Under these assumptions, the triple $(x_i,n^+,n^-)$ evolves as a continuous-time Markov chain, with the following transitions
{\footnotesize
\[
\begin{split}
(x,n^+,n^-) &\mapsto  (-x,n^+,n^-)  \text{ with rate } u(t)=\alpha(x,y,n^+,n^-,n^{\e},t);\\
(x,n^+,n^-) &\mapsto  (x,n^++1,n^-)  \text{ with rate }\gamma^+(x,n^+,n^-,n^{\e},t)=\Big(n^{\e}-n^+\Big)\!\cdot \beta\Big(\!\!-\!1,+\e,n^+\!+\mathbb{I}_{\{x=1\}},n^-,n^{\e},t\Big);\\
(x,n^+,n^-) &\mapsto  (x,n^+-1,n^-)  \text{ with rate }\delta^+(x,n^+,n^-,n^{\e},t)=n^+\cdot \beta\Big(1,+\e,n^+\!-\mathbb{I}_{\{x=-1\}},n^-,n^{\e},t\Big);\\
(x,n^+,n^-) &\mapsto  (x,n^+,n^-+1)  \text{ with rate }\gamma^-(x,n^+,n^-,n^{\e},t)=\Big(N\!-n^{\e}\!-n^-\Big)\!\cdot \beta\Big(\!\!-\!1,-\e,n^+\!,n^-\!+\mathbb{I}_{\{x=1\}},n^{\e},t\Big);\\
(x,n^+,n^-) &\mapsto  (x,n^+,n^--1)  \text{ with rate }\delta^-(x,n^+,n^-,n^{\e},t)=n^-\cdot \beta\Big(1,-\e,n^+,n^-\!-\mathbb{I}){\{x=-\!1\}},t\Big).
\end{split}
\]
}
The best response $u(t) = \alpha(x(t),y,n^+(t),n^-(t),n^{\e},t)$ for player $i$ is the one minimizing the cost
\[
\mathbb E \left[\int_0^T \frac{u^2(t)}{2} \, dt - x(T)\, (m_{N+1}(T)+y)\right],
\]
with
\[
m_{N+1}(T)=2\frac{(n^+(t)+n^-(t)) + \mathbb{I}_{\{x=1\}}}{N+1}-1.
\]
By Dynamic Programming, the Value Function for this stochastic optimal control problem solves
\[
\begin{split}
& \frac{\partial V}{\partial t}   +\min_{u} \left[\frac{u^2}{2}+u\nabla_x V  +\gamma^+ \nabla^+_{\gamma} V  + \delta^+ \nabla^+_{\delta} V  + \gamma^- \nabla^-_{\gamma} V + \delta^- \nabla^-_{\delta} V      \right]  =0 \\
& V(x,y,n^+,n^-,T)  =-x\left(\frac{2(n^++n^- + \mathbb{I}_{\{x=1\} )}}{N+1}-1\right). 
\end{split}
\]
The minimisation over $u$ leads to the optimal feedback
\begin{equation}\label{alpha}\alpha^*(x,y,n^+,n^-,t)=\left[V(-x,y,n^+,n^-,t)-V(x,y,n^+,n^-,t)\right]^-\end{equation}
and, finally, to the HJB equation
\begin{equation}\label{eq1}
\begin{split}
\dot V = & \ \frac{1}{2} \left( \left[V(-x,y,n^+,n^-,t)-V(x,y,n^+,n^-,t)\right]^-\right)^2 \\
 &  - \gamma^+(x,n^+,n^-,t)\cdot [ V(x,+\e,n^++1,n^-,t)-V(x,+\e,n^+,n^-,t) ] \\
 & - \delta^+(x,n^+,n^-,t)\cdot [ V(x,+\e,n^+-1,n^-,t)-V(x,+\e,n^+,n^-,t) ]   \\
 & - \gamma^-(x,n^+,n^-,t)\cdot [ V(x,-\e,n^+,n^-+1,t)-V(x,-\e,n^+,n^-,t) ] \\
 & - \delta^-(x,n^+,n^-,t)\cdot [ V(x,-\e,n^+,n^--1,t)-V(x,-\e,n^+,n^-,t) ].  
\end{split}
\end{equation}
The unique Nash equilibrium of the game is obtained by setting $\alpha=\beta=\alpha^*$, under which the HJB reduces to a system of $4\,(n^{\e}+1)(N-n^{\e}+1)$ ordinary differential equations in the state variables $V$, and can be solved numerically. Specifically, we use the software  Matlab  to solve an ODE system backward in time, meaning that the final conditions play the role of initial conditions and the variation is inverted in time (the r.h.s of \eqref{eq1} is multiplied by $-1$).

\subsection{Simulations of the $N$-player system}

Having obtained the Nash equilibrium feedback control $\a^*$ for $N+1$ players, we rescale the problem to $N$ players, and simulate the evolution of the sufficient statistics
\[
n^+(t) = \sum_{i=1}^N \mathbb{I}_{\{x_i = 1\}}\mathbb{I}_{\{y_i = \e\}}, \ \ n^-(t) = \sum_{i=1}^N \mathbb{I}_{\{x_i = 1\}}\mathbb{I}_{\{y_i = -\e\}},
\]
which has the Markovian evolution
\[
\begin{split}
(n^+,n^-) \mapsto  (n^++1,n^-)&\text{ with rate } \Big(n^{\e}-n^+\Big)\cdot \alpha^*(-1,+\e,n^+,n^-, n^{\e},t);\\
(n^+,n^-) \mapsto  (n^+-1,n^-)&\text{ with rate } n^+\cdot \alpha^*(1,+\e,n^+-1,n^-, n^{\e},t);\\
(n^+,n^-) \mapsto  (n^+,n^-+1)&\text{ with rate } \Big(N-n^{\e}-n^-\Big)\cdot \alpha^*(-1,-\e,n^+,n^-,n^{\e},t);\\
(n^+,n^-) \mapsto  (n^+,n^--1)&\text{ with rate } n^-\cdot \alpha^*(1,-\e,n^+,n^--1,n^{\e},t).
\end{split}
\]
Initializing appropriately $(n^+(0),n^-(0))$, by independently assigning to each player $x_i(0)=1$ with probability $\frac{1+m_0}{2}$ and $y_i=\e$ with probability $\frac12$, we run simulations of the above dynamics, eventually obtaining samples for $m_N(T) = 2\frac{n^+(T)+n^-(T)}{N}-1$. Averaging over $S$ independent simulations we estimate the expectation $\mathbb E(m_N(T))$.

More in detail, in our first series of experiments we fix $m_0=0.25$, and we take different values of $\e$ and $T$. Concerning the number of simulations, we set $S=100$, whereas the number of agents in the population is $N=30$. This figure could appear too small to describe a \emph{large population}. However, what we see in our simulations is that the expected values of $m_N$ are approximating very well the asymptotic equilibrium of the mean field game, except for some \emph{transition windows} that we will discuss in more detail. Later, in a second series of experiments, we will also consider $N=60$. Note that, by increasing $N$, the numerical problem becomes quickly intractable because of the high dimension of the HJB associated with the $N$-dimensional system.\footnote{For $N=30$, we have a system of 1024 equations. For $N=60$, the number increases to 3844.}

\begin{figure}[h]
 \centering
\includegraphics[scale=0.45]{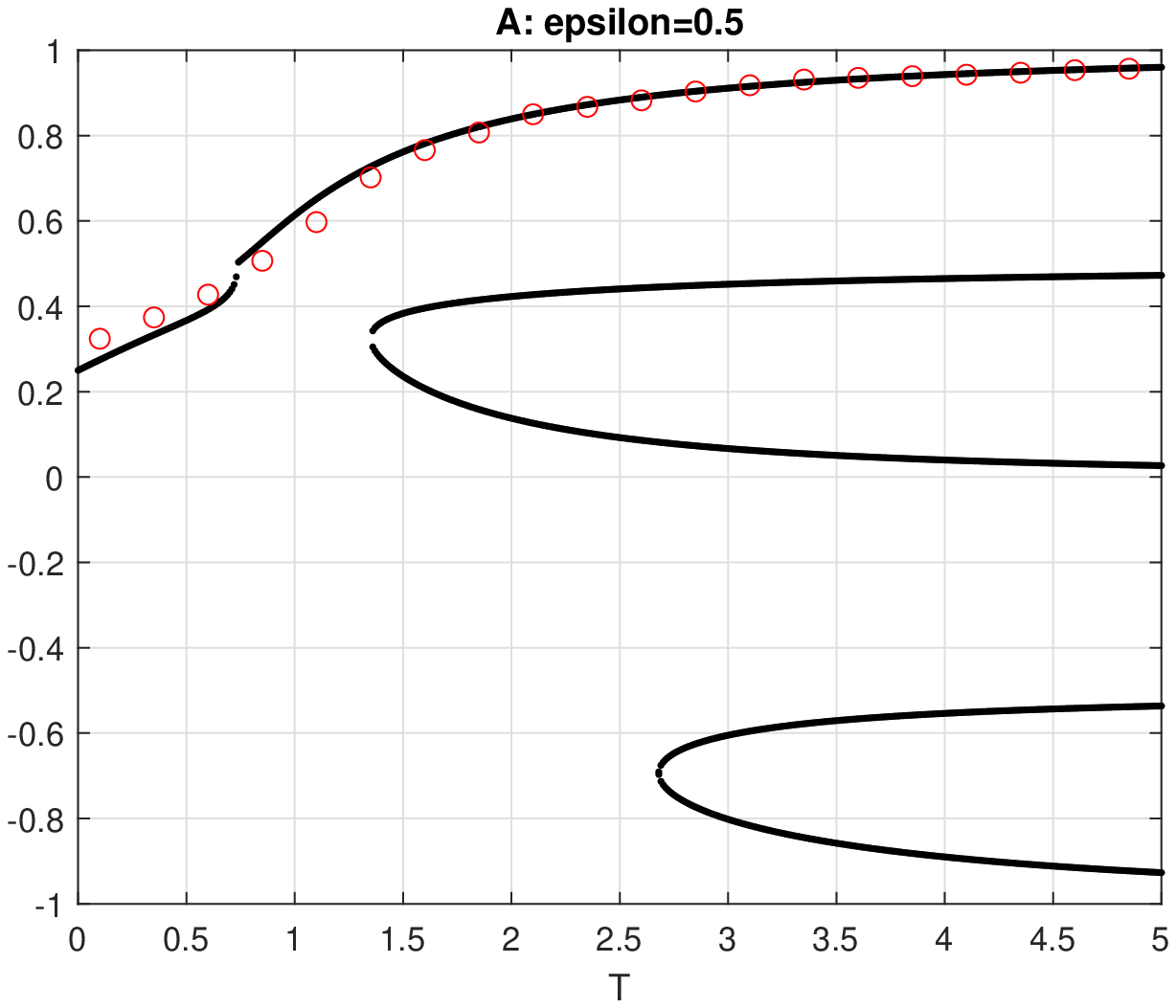} \\
\includegraphics[scale=0.34]{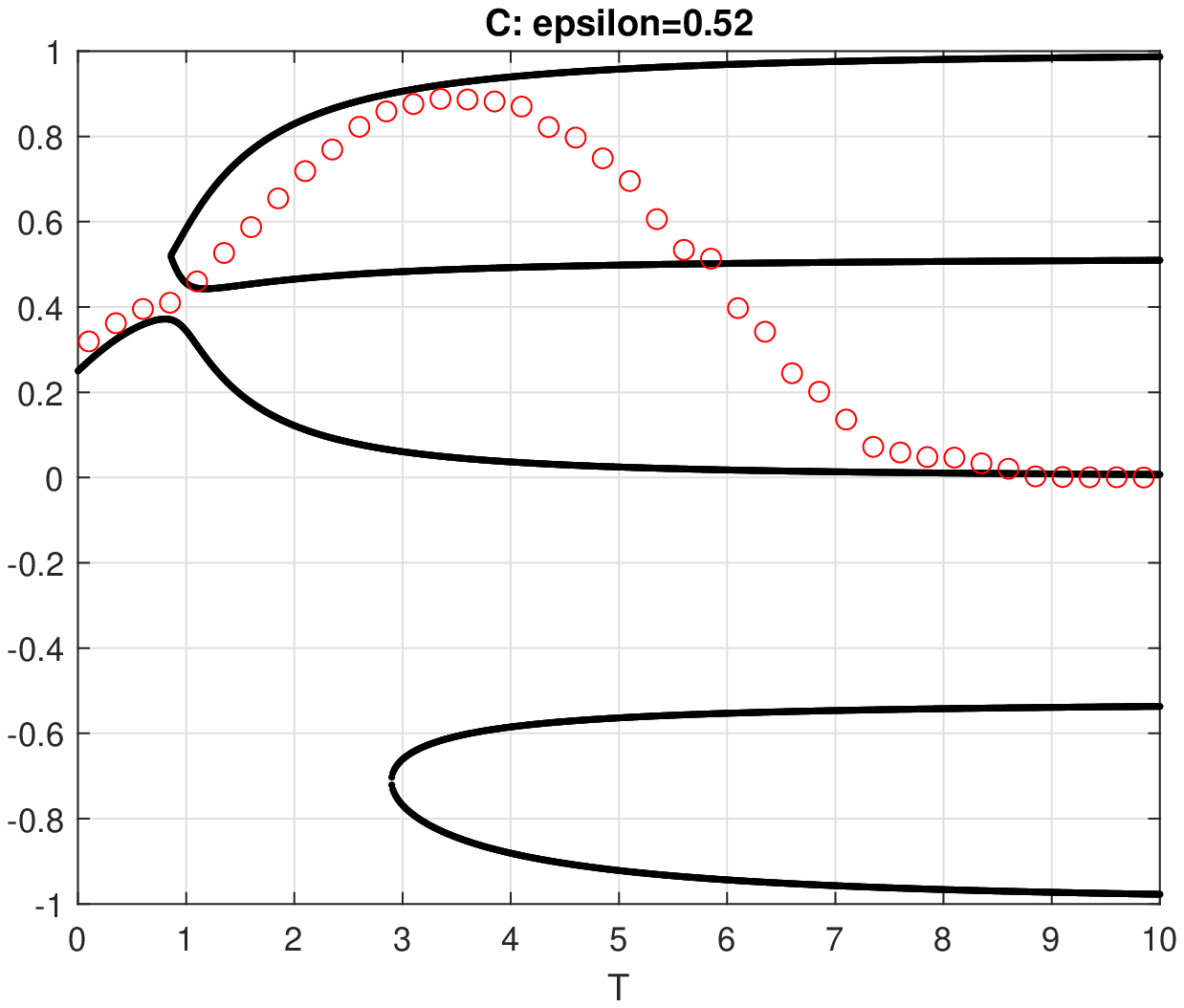}
\includegraphics[scale=0.33]{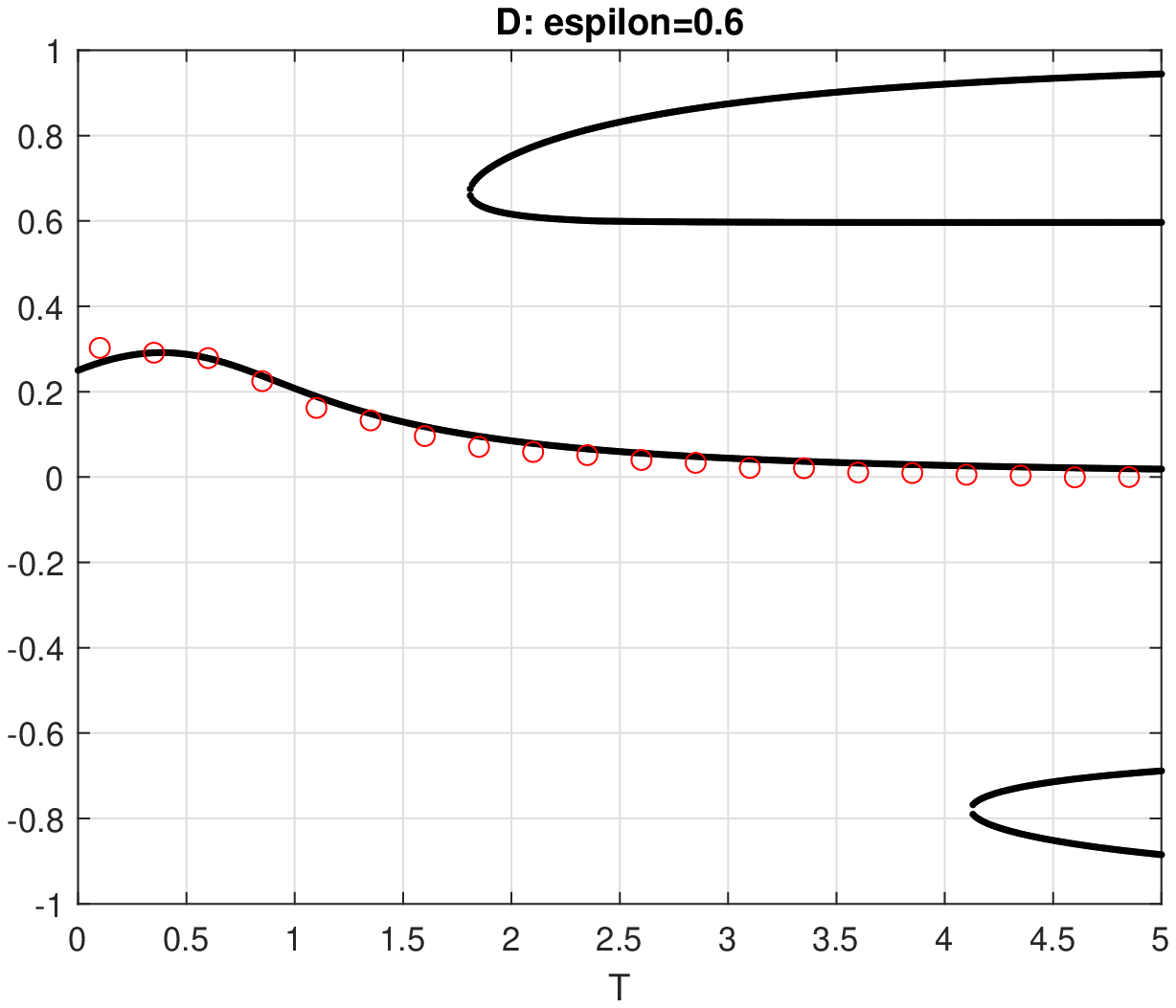}
\vspace{-.1 in}
\caption{Values of $m(T)$ (black dots) and $m_N(T)$ (red large circles).}
\label{fig2}
\end{figure}

In Figure \ref{fig2}, we   compare $\mathbb{E}(m_N(T))$ (red circles), the equilibrium emerging in the finite dimensional model, with the mean field equilibria described as the solutions to the Fixed Point Equation \eqref{eq} (black lines). Specifically, we consider four different values of $\e\in\{0.5, \ 0.52,\ 0.6\}$ and we let $T$ vary from $T=0$ to a large time, where the largest equilibrium value of $m(T)$ is approaching the limit value of $1$. We expect that $\mathbb{E}(m_N(T))$ converges in $N$ towards one of the mean field equilibria solving \eqref{eq}. This is verified in our simulations; notably,  for certain values of the parameter $\e$, we see a clear transition from a polarized to a unpolarized equilibrium or vice versa. In fact, by looking at the four panels of Figure \ref{fig2}, if $\e$ is large enough (see Panel C), the individual behavior prevails for all $T$ and the population sticks with the smallest (unpolarized and coherent) equilibrium. When $\e$ is small enough (see Panel A), the selected equilibrium changes continuously in $T$, as in the large $\e$ case, but this equilibrium is unpolarized for small $T$ and polarized for larger $T$. More interesting is the  case of intermediate values of $\e$ (see Panel B). Here we see a continuous branch of unpolarized and coherent equilibria existing for all $T>0$, while a branch of polarized coherent equilibria emerges for $T$ sufficiently large. In this case the $N$-player game agrees with the unique unpolarized equilibrium for small $T$, jumps to the branch of polarized equilibria as it emerges, but for larger $T$ it jumps back to the less polarized equilibrium. We actually see ``smooth transitions'' rather than ``jumps'', but this could be due to the small value of $N$ in simulations.

\begin{figure}[h]
 \centering
\includegraphics[scale=0.45]{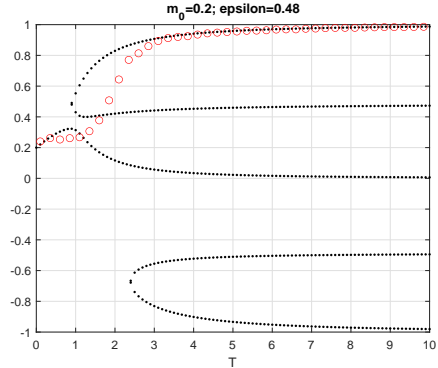}
\vspace{-.1 in}
\caption{Values of $m(T)$ (black dots) and $m_N(T)$ (red large circles). Here, $m_0=0.2$ and $\e=0.48$.}
\label{fig3}
\end{figure}

Note that this switch from polarized to unpolarized is not seen for all values of the initial condition $m_0$, as seen in Figure  \ref{fig3}.

In the next section, we provide a justification behind the emergence of one selected equilibrium, in case that multiple equilibria are present in the mean field limit.

\subsection{A rationale behind the  equilibrium selection}\label{sec:N_economic}

In all the numerical experiments we have performed about $m_N(T)$, the equilibrium emerging in the $N$-dimensional system, we can recognize two important properties:
\vspace{.3cm}
\\
\underline{Property 1.} The equilibrium $m_N(T)$ is always coherent. Namely, if $m_0>0$, then $m_N(T)>0$. 
\vspace{.1cm}
\\
\underline{Property 2.} For some values of $\e$,  $m_N(T)$ switches from a polarized to an unpolarized equilibrium (or vice versa) depending on the length of the time horizon $T$. 
\vspace{.3cm}
\\
Concerning Property 1,  when we look at the finite dimensional system, the equilibrium $m_N(T)$ converges, for $N$ large, to one of the values $m(T)$  solving $\eqref{eq}$. Note that, among those values, there is at least one coherent equilibrium (i.e., an equilibrium with the same sign as $m_0$).  It is then plausible to presume that the finite population will select one of the coherent equilibria, in that conveying to an incoherent equilibrium would ask for a (implausible) mobilization of the subpopulation ex-ante aligned with the sign of $m_0$. Property 2., instead, deals with the eventual polarization of the coherent  equilibrium selected when playing the finite dimensional game. Here the discussion is more subtle, since we do not have a clear and trivial explanation of the evident phase transitions we see in the simulations. The simplest explanation could be that the population chooses the equilibrium which minimizes the \emph{total cost},  $J(u)$, defined in \eqref{costmfg}. Interestingly, this functional, being a function of the control, can be rewritten in terms of $m(T)$. We can take advantage of \eqref{eq:v} to see that, depending on the value of $x\in\{-1,+1\}$ and $y\in\{-\e,+\e\}$, we can specify the cost needed to reach a certain equilibrium value $m(T)$: 
\begin{equation}\label{eq:v_new}
 v_{m(T)}(x,y) = \left\{ \begin{array}{ll} -|m(T)+y| &\quad \mbox{ if } \sign(m(T)+y) \in \{0,x\} \\ -|m(T)+y| + \frac{2|m(T)+y|}{T\,|m(T)+y|+1} & \quad \mbox{ if } \sign(m(T)+y) = -x,\end{array} \right.
 \end{equation}
 where $v_{m(T)}(x,y)=V(0,x,y)$ evaluated at $m=m(T)$.
This functional describes  the total cost sustained by each subpopulation indexed by $x$ and $y$ to reach the equilibrium $m(T)$. We can now derive the costs sustained by  the \emph{underdog} subpopulation (the one whose local filed is opposite in sign to $m_0$) and by its opponent, namely the one whose local filed is aligned with $m_0$. 
With a slight abuse of notation, we denote such quantities with $J^{(-\e)}(m(T))$ and $J^{(+\e)}(m(T))$ to emphasize the dependence on the prevailing equilibrium. Accordingly, we will also write $J(m(T))$ as the total cost sustained by the entire system. We have
$$J^{(-\e)}(m(T))=\frac{1-m_0}{2}\, v_{m(T)}(-1,-\e)+\frac{1+m_0}{2}\, v_{m(T)}(+1,-\e)$$
$$J^{(+\e)}(m(T))=\frac{1-m_0}{2}\, v_{m(T)}(-1,+\e)+\frac{1+m_0}{2}\, v_{m(T)}(+1,+\e)$$
and
$$J(m(T))=\frac{1}{2} J^{(-\e)}(m(T))+\frac{1}{2} J^{(-\e)}(m(T)).$$
It is not difficult to  see that, when considering only coherent equilibria (i.e., $m(T)$ such that $\sign(m(T))=\sign(m_0)$),  $J(m(T))$ decreases in $m(T)$, in the sense that the more polarized the equilibrium is, the lower the cost to reach it. Therefore, we could expect the polarized equilibrium to prevail. However, as said, for certain values of the parameters, this is not the case. In line with our simulations, we now shape a different  conjecture. We see that the prevailing equilibrium is the one that, among the coherent ones, minimizes the functional $J^{(-\e)}$, namely the cost related to the \emph{underdog} subpopulation. We rephrase this conjecture in the following fact, which embraces both the previous properties.
\vspace{.3cm}
\\
\underline{Property 3.} The equilibrium $m_N(T)$ of the $N$-dimensional system converges to the coherent solution of \eqref{eq} that minimizes $J^{(-\e)}$.

In some sense, abstracting a two-player game played between the favorite player ($y=+\e$) and the underdog one ($y=-\e$), we can say that the former \emph{imposes}  that the equilibrium will be coherent (and this minimizes her effort), whereas the latter decides about polarization (again, minimizing effort given the previous selection). 

\begin{figure}[h]
 \centering
 \includegraphics[scale=.55]{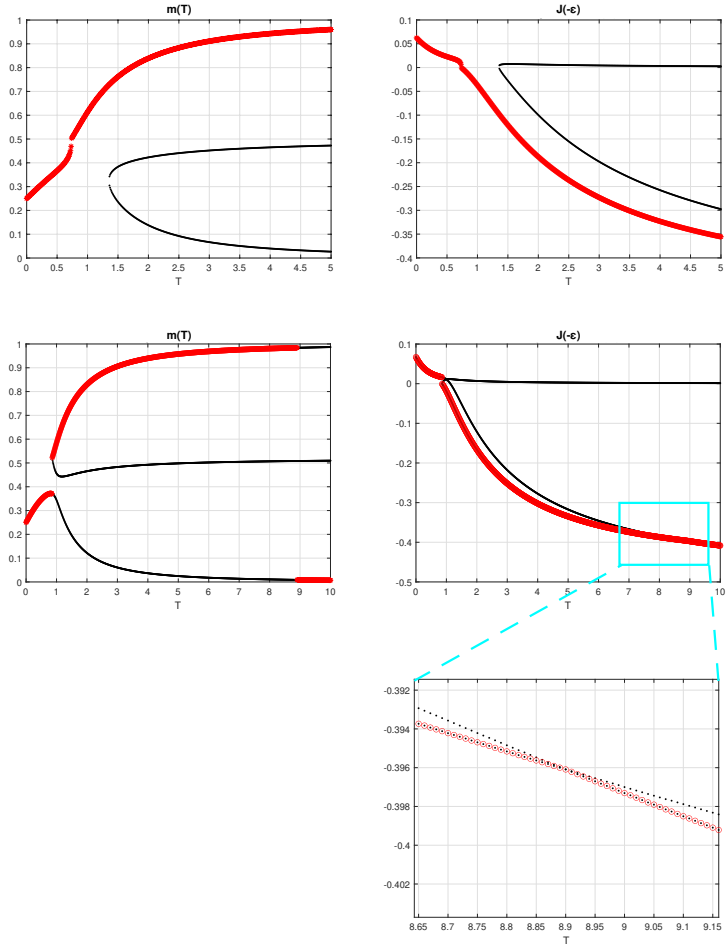}
\vspace{-.1 in}
\caption{Phase diagram for $m(T)$ (left panels) and $J^{(-\e)}$ (right panels) for two different values of $\e$. Here, $\e=0.5$ (top panels) and $\e=0.52$ (bottom panels). In the left panels, the red points denote the solution $m(T)$ corresponding to the minimum value of $J^{(-\e)}$, (depicted in red in the corresponding right panel). }
\label{fig4}
\end{figure}

To provide evidence about the goodness of Property 3., in Figure \ref{fig4}, we plot the phase diagram of $J^{(-\e)}$ for $m_0=0.25$ and for the same values of $\e$ and $T$ seen in Figure \ref{fig2}. As said before, for each equilibrium value $m(T)$ of the mean field limit, we have the corresponding value of  $J^{(-\e)}$. 

Note that, for $\e=0.52$, we see two transitions corresponding to the points where the two branches of  $J^{(-\e)}$ related to the polarized and unpolarized coherent equilibria intersect themselves. In the lower panel of Figure \ref{fig4}, we zoom on the right-bottom panel to better recognize such intersection. We see that the two branches of  $J^{(-\e)}$ related to the polarized and unpolarized coherent equilibria intersect themselves at $T\approx 8.9$. Notably, this point lies in the time interval, where the emerging equilibrium of the $N$-dimensional system  jumps from the polarized equilibrium to the unpolarized one (see Panel C of Figure \ref{fig2}). We do not report all the figures related to the other values of $\e$, but the same fact still appears, thus corroborating Property 3.

\begin{figure}[h]
 \centering
\includegraphics[scale=.5]{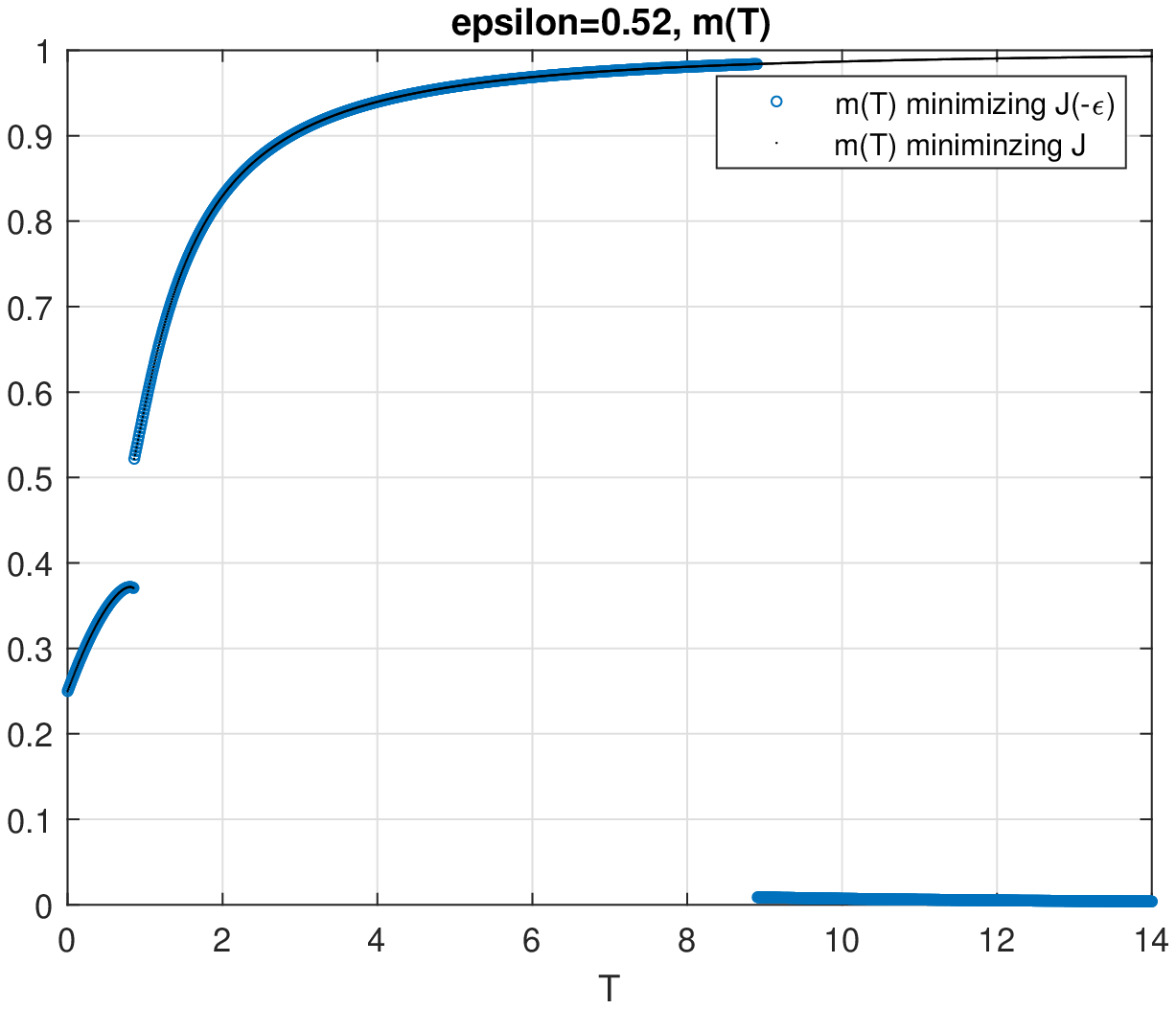} 
\includegraphics[scale=.5]{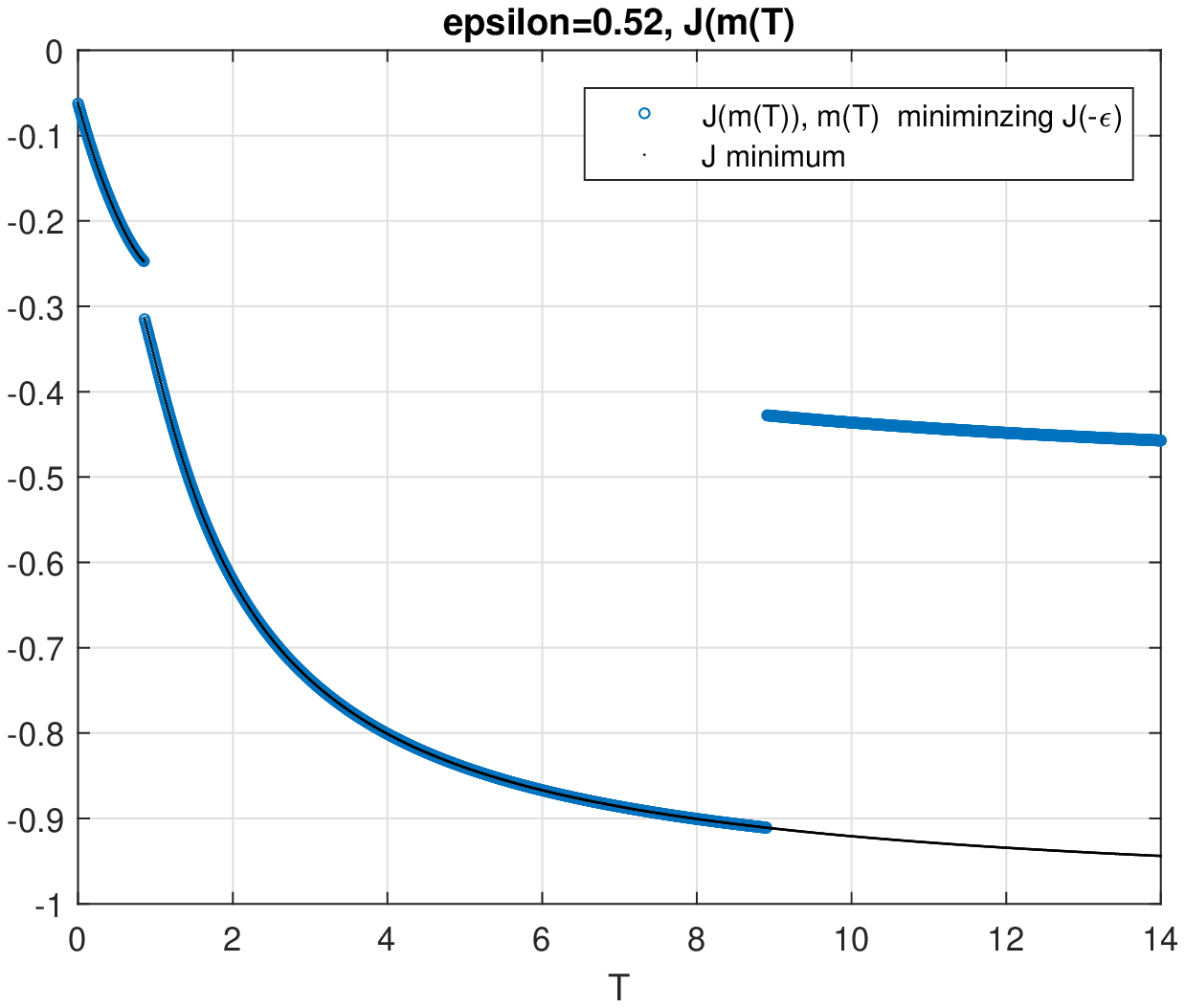}
\caption{Left: The value of $m(T)$ minimizing the functional $J^{(-\e)}$ (bold blue circles) and minimizing $J$ (thin black line). Right: The functional $J$ computed for $m(T)$ minimizing  $J^{(-\e)}$ (bold circles) and the minimum level of $J$ (thin black line). Here, $\e=0.52$. }
\vspace{-.1 in}
\label{fig6}
\end{figure}

Finally, we show that the solution $m(T)$ related to the prevailing solution $m_N(T)$ does not necessarily minimize to total cost $J(m(T))$. In Figure \ref{fig6} (left panel), for $\e=0.52$, we plot the value of $m(T)$ that minimizes $J^{(-\e)}$ (bold blue circles) and $J$ (thin black line). It can be seen that the two cases coincide as soon as $T$ exceed the value $T\approx 8.9$ discussed above. In the right panel of the same figure, we plot two different branches of $J$, one corresponding to $m(T)$ which minimizes $J^{(-\e)}$, and the second one related to $m(T)$ which minimizes $J$ (thin black line). We see that the two curves differ exactly for $T$ larger than the intersection value depicted above; this is the value of $T$  where we know that the equilibrium $m_N(T)$ in the finite dimensional model jumps from the polarized to the unpolarized one. This shows that the equilibrium emerging in the population dynamics does not always minimize the total cost. In some sense,  the unpolarized equilibrium  partially  favors the underdog subpopulation, in that $J^{(-\e)}$ is minimized among the coherent equilibria. 

We now run a second series of experiment; the aim is still to reinforce the goodness of Property 3, showing that $m_N(T)$ fairly well approximates the coherent mean field game equilibrium minimizing $J^{(-\e)}$. For this round of simulations, we increase the number of agents taking $N=60$. Concerning the other parameters, we choose different values of  $m_0$, $\e$  and $T$, in order to consider cases where there are one, three or five solutions to \eqref{eq}. Specifically, in the first panel of simulations, we fix $T=1$ and let $m_0$ and $\e$ vary (cf. the first six experiments reported in Table \ref{tab2}); in the second panel, we fix $m_0=0.2$ and let $T$ and $\e$ vary (cf. the second six experiments reported in Table \ref{tab2}).  In Table \ref{tab2}, we report all results. The first three columns summarize the parameters of each experiment.  In the fourth, we report the value of $m(T)$ which matches what prescribed by Property 3. Columns five and six pertain to the finite dimensional model; we indicate by $\overline{m}_N(T)$  the average value of $m_N(T)$ as resulting by running $S=100$ simulations, with an indication of its standard deviation,  $SD(m_N(T))$. Finally, in columns seven and eight, we measure the goodness of the approximation by reporting $|\overline{m}_N(T)-m(T)|$ and $|\overline{m}_N(T)-m(T)|/SD(m_N(T))$, respectively. By looking at these latter columns, it is evident that the equilibrium prevailing in the finite dimensional model aligns with what prescribed by Property 3. This is testified by the fact that the difference between $\overline{m}_N(T)$ and $m(T)$ is close to $0$ (the highest value is $0.038$ in experiment n.7) and that such difference is always below one standard deviation (the highest ratio is, again, in experiment n.7).

 \begin{table}[ht]
  \begin{center}
  \begin{tabular}[]{ccc|c|cc|cc}
 $T$ &$m_0$ & $\e$ & $m(T)$ &  $\overline{m}_N(T)$ & $SD(m_N(T))$ & $|\overline{m}_N(T)-m(T)|$  & $\frac{|\overline{m}_N(T)-m(T)|}{SD(m_N(T))}$\\
 \hline\hline
1&       0.1 &	0.42 & 0.8261 & 0.8240 & 0.0794 & 0.0021	& 0.0264\\
1&    0.1 & 0.45 & 0.8126 & 0.8020 & 0.0859 & 0.0106	 & 0.1234\\
1&    0.1 & 0.5 & 0.0506 & 0.0390 & 0.0765&  0.0116&0.1516\\
1& 0.5 & 0.55 & 0.8962 & 0.9171 & 0.0553& 0.0209&0.3779\\
1& 0.5 & 0.6 & 0.8818 & 0.8860 & 0.0569&  0.0042&0.0738\\
1&  0.5 & 0.7 & 0.1276 & 0.1080 & 0.0783&  0.0196&0.2503\\
\hline
2.3&    0.2 &	0.5 & 0.8552 & 0.8173 & 0.0788&  0.0379 & 0.4810\\
2.3&    0.2 &	0.58 & 0.0583 & 0.0460 & 0.0818& 0.0123&0.1504\\
2.8&    0.2 &	0.5 & 0.8925 & 0.8827 & 0.0715&  0.0098&0.1371\\
3.5&    0.2 &	0.7 & 0.0203 & 0.0173 & 0.0473&  0.0030&0.0634\\
5.5&    0.2 &	0.7 & 0.0094 & 0.0077 & 0.0307&  0.0017&0.0554\\
9&    0.2 &	0.7 & 0.0039 & 0.0030 & 0.0171& 0.0009&0.0526\\
	  \end{tabular}
  \vspace{.1 in}
	    \caption{Results of numerical experiments run to check the goodness of Property 3.} \label{tab2}
	  \end{center}
\end{table}

\section{Conclusions}\label{sec:conclusion}
We have studied a simple mean field game on a time interval $[0,T]$, where players can control their binary state according to a functional made of a quadratic cost and a final reward. This latter depends on two competing drivers: (i) a social component rewarding conformism, namely,  being part of  the majority  (conformism) and (ii) a private signal favoring the coherence of individuals with respect to a personal type (stubbornness). The trade-off between these two factors, associated with the antimonotonicity of the objective functional,  leads to a fairly reach phase diagram. Specifically, the presence of multiple Nash Equilibria for the mean field game has been detected; moreover,  when looking at the aggregate outcome of the game,  several different types of equilibria can emerge in terms of polarization (fraction of conformists) and coherence (sign of the majority at the final time $T$ compared to the sign of the initial condition). 

We have described and characterized the full phase diagram and discussed the role of all the parameters of the model with respect to the aforementioned classification of possible equilibria. We have also analyzed a  $N$-player version of the same mean field game. It is a well-known that in this latter case, the Nash Equilibrium is necessarily unique. It becomes, therefore, interesting to identify which equilibrium is selected by the $N$-finite population, in case the corresponding mean field game exhibits multiple equilibria. In this respect, we detected phase transitions, in the sense that, depending on one or more parameters of the model, the equilibrium emerging in the finite-dimensional game is always coherent, but it  may turn from an unpolarized one to a polarized and vice versa, depending on the length of the time horizon, $T$.  This fact seems to be new in the mean field game literature. At a first glance, we could expect the finite dimensional population to select the equilibrium minimizing the cost associated to the equilibrium for the entire system (interpreted as the collective cost). By contrast, what emerges from our simulations is that the equilibrium prevailing in the finite dimensional game is the one converging, for $N$ large, to the coherent equilibrium that minimizes the cost functional associated with the \emph{ex-ante underdog} subpopulation, namely, the collection of such players whose private signal opposites  the sign of the majority at time zero. Put differently, it seems that the \emph{ex-ante} favorite subpopulation (namely, the one whose private signal is aligned with the initial condition) imposes the selection of a coherent equilibrium, whereas the \emph{ex-ante} underdog subpopulation (namely, the one whose private signal opposites the sign of the initial condition) decides about polarization.

\appendix  
\section{Proof of results}

\subsection*{Proof of Proposition \ref{prop:equilibria}: polarized coherent equilibria ($m > \e$)} 
For $m>\e$ \eqref{eq} can be rewritten as
\be{eq2}
F(m,\e,T,m_0) :=1-m - \frac{1-m_0}{\e^2 T^2} \varphi\left(\frac{1+mT}{\e\, T} \right) = 0,
\ee
where
\[
\varphi(y) = \frac{y^2 + 1}{(y^2-1)^2}.
\]
Before describing the solutions of the equation $F(m,\e,T,m_0) $ in $(\e,1]$ we give some simple facts that will be useful in the proof. 

\noindent{\bf Fact 1}. The map $m \mapsto F(m,\e,T,m_0)$ is strictly concave for $m \in (\e,1]$, for each $\e,T,m_0$.

{\it Proof.}
This comes immediately from the fact that $\varphi$ is strictly convex in $(1,+\infty)$. Indeed,
\be{derphi}
\varphi'(y) = - \frac{2y(3+y^2)}{(y^2-1)^3}, \ \ \ \varphi''(y) = \frac{6(y^4 + 6 y^2 + 1)}{(y^2-1)^4}.
\ee
Note also that $\varphi$ is strictly decreasing.\qed

As a consequence of Fact 1, \eqref{eq2} can have at most two solutions in $(\e,1]$.

\noindent{\bf Fact 2}. The map $T \mapsto F(m,\e,T,m_0)$ is strictly increasing for $T \in [0,+\infty)$, for each $m,\e,m_0$, with $m > \e$.

{\it Proof.} To see this, note that 
\[
\frac{\partial F}{\partial T} (m,\e,T,m_0) = \frac{2(1-m_0)\left[m(1+Tm)^3 - 3 \e^2 T(1+Tm) - \e^4 T^3\right]}{\left((1+Tm)^3 - \e^2 T^2\right)^3},
\]
which has the same sign as
 $
    h(m,\e,T,m_0) := m(1+Tm)^3 - 3 \e^2 T(1+Tm) - \e^4 T^3.
$ 
Observing that $h(\e,\e,T,m_0) = \e$, that 
\[
\frac{\partial h}{\partial m} (m,\e,T,m_0) = (1+Tm)^3 + 3 Tm(1+Tm)^2 - 3 \e^2 T^2
\]
is increasing in $m$ and 
 $
\frac{\partial h}{\partial m} (\e,\e,T,m_0)  = 4 \e^3 T^3 + 6 \e^2 T^2 + 6 \e\, T + 1 >0,
$ 
we conclude that $\frac{\partial F}{\partial T} (m,\e,T,m_0)>0$ for $m \in (\e,1]$. This proves Fact2.\qed

\noindent{\bf Fact 3}. $\forall m,\e,m_0$, with $m > \e,$
\[
F(m,\e,0^+,m_0):=\lim_{T \downarrow 0} F(m,\e,T,m_0) = m_0 - m
\]
and 
\[
 \lim_{T \uparrow +\infty} F(\e,\e,T,m_0) = \frac{1+m_0-2\e}{2},
\]
whereas for $m \in (\e,1]$,
\[
 \lim_{T \uparrow +\infty} F(\e,\e,T,m_0) = 1-m.
\]

These are simple asymptotics, and the proof is omitted.

Now we can prove Proposition \ref{prop:equilibria}. We begin with the case $\e \leq m_0$. By Fact 3, 
 $
F(\e,\e,0^+,m_0) = m_0 - \e \geq 0;
$ 
thus, by Fact 2, $F(\e,\e,T,m_0)>0$, $\forall T>0$. Since, clearly, $F(1,\e,T,m_0)<0$ and $m \mapsto F(m,\e,T,m_0)$ is strictly concave by Fact 1, the uniqueness of the solution to \eqref{eq2} follows readily and case (i) is proved.

Now, consider the case  $\e \geq \frac{1+m_0}{2}$ (which implies $\e>m_0$) . By Facts 2 and 3, $F(\e,\e,T,m_0)< 0$, $\forall T>0$. 
Since $F(m,\e,0^+,m_0) = m_0 - m$, by continuity \eqref{eq2} has no solution in $(\e,1]$ for $T$ small enough. Now we claim that there exists a unique $T_c^{(1)} = T_c^{(1)}(\e,m_0)>0$ such that the graph of $z= F(m,\e,T_c^{(1)},m_0)$ is tangent to the line $z=0$. Since $F(m,\e,T,m_0)$ is strictly increasing in $T$ (Fact 2), such $T_c^{(1)}$, if any,  is unique. To prove the existence of $T_c^{(1)}$ by continuity it is enough to show that there are values of $m \in (\e,1]$ and $T>0$ such that $F(m,\e,T,m_0)>0$: this is true $\forall m \in (\e,1)$ as $F(m,\e,T,m_0) \ra 1-m$ for $T \uparrow +\infty$ (Fact 3). The conclusions in case (ii) are now obvious consequences of concavity and $T$-monotonicity of $F$.

Consider, finally, the case $m_0 < \e < \frac{1+m_0}{2}$.
We first investigate the following equation in the unknown $T>0$:
\be{eq0}
F(\e,\e,T,m_0)=0,
\ee
that can be rewritten as 
\[
1-\e - \frac{2\e^2 T^2 + 2 \e \,T + 1}{4 \e^2 T^2 + 4 \e \,T + 1} (1-m_0) = 0.
\]
This can be cast as a quadratic equation in $T$: it has real solutions if and only if $\e < \frac{1+m_0}{2}$, and in this case the only positive solution is $T^*(\e,m_0)$ given in \eqref{Tstar}. Note that $\e > m_0$ implies that $T^*(\e,m_0)>0$. Now, $T$-monotonicity and $m$-concavity of $F(m,\e,T,m_0)$ and the fact that
\be{dere}
\frac{\partial}{\partial m} F(\e,\e,T,m_0) = -1+2(1-m_0) \frac{T(\e \,T+1)(4 \e^2 T^2 + 2 \e\, T + 1)}{2\e\, T + 1}
\ee
is strictly increasing in $T$,
show that there are two alternatives:

\ben
\item
if $\frac{\partial}{\partial m} F(\e,\e,T^*(\e,m_0),m_0) \leq 0$, then, 
\bi
\item
for $T \leq T^*(\e,m_0)$, the map $m \mapsto F(m,\e,T,m_0)$ is less than or equal to $0$ at $m = \e$ and it is decreasing: there are no solutions to \eqref{eq2} in $(\e,1]$;
\item
for $T > T^*(\e,m_0)$, the map $m \mapsto F(m,\e,T,m_0)$ is strictly positive at $m = \e$, concave and negative at $m=1$, so \eqref{eq2} has a unique solution in $(\e,1]$;
\ei

\item
if $\frac{\partial}{\partial m} F(\e,\e,T^*(\e,m_0),m_0) > 0$, then by continuity there exists $T_c^{(1)} = T_c^{(1)}(\e,m_0) \in (0,T^*(\e,m_0))$ such that the graph of $z= F(m,\e,T_c,m_0)$ is tangent to the line $z=0$; as above, this $T_c^{(1)}$ is unique. Thus,
\bi
\item
for $T<T_c^{(1)}(\e,m_0)$, \eqref{eq2} has no solutions in $(\e,1]$;
\item
for $T=T_c^{(1)}(\e,m_0)$, \eqref{eq2} has a unique solution in $(\e,1]$;
\item
for $T_c^{(1)}(\e,m_0)<T<T^*(\e, m_0)$, \eqref{eq2} has two solutions in $(\e,1]$;
\item
for $T \geq T^*(\e, m_0)$, \eqref{eq2} has a unique solution in $(\e,1]$.
\ei
\een
Note that $T_c^{(1)}$ is defined as in case (ii), and by the Implicit Function Theorem it is continuous at $\e$, $\forall\e$ in its domain. To complete the proof of case (iii) we are left to show that there exists $\e_*^{(1)} \in \left(m_0,\frac{1+m_0}{2}\right)$ such that 
\[
\frac{\partial}{\partial m} F(\e,\e,T^*(\e,m_0),m_0) > 0 \ \  \mbox{ if and only if } \ \  \e \in \left(\e_*^{(1)},\frac{1+m_0}{2}\right), 
\]
so that $T_c^{(1)}(\e,m_0)$ is defined for $\e \in (\e_*^{(1)},1]$. Moreover, if this is the case, continuity implies that 
\[
\lim_{\e \downarrow \e_*^{(1)}} T_c^{(1)}(\e,m_0)= T^*(\e^*,m_0).
\]
To complete the proof we are left to show the existence of such $\e_*^{(1)}$. This is established by proving that the map
 $
\e \mapsto \frac{\partial}{\partial m} F(\e,\e,T^*(\e,m_0),m_0)
$ 
is strictly increasing, negative at $\e = m_0$ and diverging to $+\infty$ at $\e = \frac{1+m_0}{2}$. We use the expressions \eqref{dere} and \eqref{Tstar}, and the change of variable 
 $
y := 1 + 2 \e\, T^*(\e).
$ 
Note that, by \eqref{Tstar},
\[
\e \,T^*(\e) =  - \frac{1}{2} + \frac{1}{2} \sqrt{\frac{1-m_0}{1+m_0-2\e}},
\]
so $y = \sqrt{\frac{1-m_0}{1+m_0-2\e}}$. It is easily seen that $\frac{dy}{d\e} > 0$, $y(m_0) = 1$ and
 $
\lim_{\e \uparrow \frac{1+m_0}{2}} y(\e) = +\infty.
$
\\
We can write, using \eqref{dere},
\[
G(\e,m_0) := \frac{\partial}{\partial m} F(\e,\e,T^*(\e,m_0),m_0) = -1 + (1-m_0) \frac{(y^2-1)(y^2-y+1)}{2 \e y}.
\]
This shows that $G(m_0,m_0) = -1$ and that $G(\e,m_0)$ diverges to $+\infty$ as $\e \uparrow \frac{1+m_0}{2}$. So the last step is to prove that $G$ is strictly increasing in $\e$. We have
\[
\frac{\partial G}{\partial \e}(\e,m_0) = \frac{1-m_0}{2\e} \frac{d}{dy} \left( \frac{(y^2-1)(y^2-y+1)}{y} \right) \frac{dy}{d\e} - \frac{1-m_0}{2 \e^2} \frac{(y^2-1)(y^2-y+1)}{y}.
\]
Using the facts that 
$
\frac{dy}{d\e} = \frac{y^3}{1-m_0} \mbox{ and } \e = \frac{(1+m_0)y^2 - (1-m_0)}{y^2},
$ 
it follows that $\frac{\partial G}{\partial \e}(\e,m_0) $ has the same sign as
\begin{multline} \label{eq3}
\frac{\e y^3}{1-m_0} \frac{3 y^4 - 2 y^3 - y^2 + 1}{y^2} -  \frac{(y^2-1)(y^2-y+1)}{y} \\ = \frac{\frac{1+m_0}{1-m_0} y^2 - 1}{y}(y^2(y-1)(3y+1) + 1) -  \frac{(y^2-1)(y^2-y+1)}{y} \\  \geq \frac{ y^2 - 1}{y}(y^2(y-1)(3y+1) + 1) -  \frac{(y^2-1)(y^2-y+1)}{y} \\
 = (y^2 - 1)(y-1)(3y^2 +y - 1) > 0, \ \ \forall \, y>1,
\end{multline}
where we have used the fact that $m_0 \geq 0$ and the expression in the second line of \eqref{eq3} is increasing in $m_0$. This completes the proof.
\qed

\subsection*{Proof of Proposition \ref{prop:equilibria2}: polarized incoherent equilibria ($m < -\e$)}
The proof repeats some of the arguments seen in the proof of Proposition \ref{prop:equilibria}. It is convenient to take advantage of the symmetry relation \eqref{simmF}. 
This implies that we can equivalently find the equilibria in $(\e,1]$ after replacing $m_0$ by $-m_0$. 
For the regime $\e\geq\frac{1-m_0}{2}$ the proof of part (ii) of Proposition \ref{prop:equilibria} applies with no changes, as the assumption $m_0 \geq 0$ was not used. \\For the case $0 < \e < \frac{1-m_0}{2}$ we can adapt the proof of part (iii) of Proposition \ref{prop:equilibria}, where the assumption $m_0 \geq 0$ was only used to prove the existence of $\e_*^{(1)}$.  Here we obtain the same behavior seen for $\e_*^{(1)} < \e < \frac{1+m_0}{2}$ in Proposition \ref{prop:equilibria}: to repeat the same argument we need to show that
\[
\frac{\partial}{\partial m} F(\e,\e,T^*(\e,-m_0),-m_0) > 0, \ \ \ \forall\e: \ \ 0 < \e < \frac{1-m_0}{2}.
\]
Indeed, as seen in the proof of Proposition \ref{prop:equilibria}, 
\[
\frac{\partial}{\partial m} F(\e,\e,T^*(\e,-m_0),-m_0) = -1 + (1+m_0) \frac{(y^2-1)(y^2-y+1)}{2 \e y},
\]
with  $y = \sqrt{\frac{1+m_0}{1-m_0-2\e}}$. Note that, as $m_0  \geq 0$, $y >1$ for all $ 0 < \e < \frac{1+m_0}{2}$. In particular, $y^2-y+1>1$ so that, with a further simple computation we get
\[
\frac{\partial}{\partial m} F(\e,\e,T^*(\e,-m_0),-m_0)> -1+ (1+m_0) \frac{y^2-1}{2 \e y} = -1 + \frac{\e + m_0}{\e}y >0.
\]
Thus, the proof for the case $0 < \e < \frac{1-m_0}{2}$  can be carried out in the same way as the proof of part (iii) of Proposition \ref{prop:equilibria} (case $\e_*^{(1)} < \e < \frac{1+m_0}{2}$).
\qed

\subsection*{Proof of Proposition \ref{prop:equilibria3}: unpolarized coherent equilibria ($0\leq m\leq\e$)}
For $m\in [0,\e]$, equation   \eqref{eq} becomes:
\be{eq2alt}
F(m,\e,T,m_0) := \frac{2(\e \,T+1)Tm + m_0 \left[ (\e\, T+1)^2 + m^2 T^2 \right]}{\left((\e\, T+1)^2 - m^2 T^2\right)^2} - m = 0.
\ee
We begin by observing that
\be{reord}
F(m,\e,T,m_0) = \frac{1}{(\e\, T+1)^2} \varphi\left(\frac{T}{1+\e\, T} m, m_0\right) - m,
\ee
where
\be{phifun}
\varphi(z,m_0) := \frac{2z+m_0 (1+z^2)}{(1-z^2)^2}.
\ee
In particular, this allows to prove easily that, if $m_0 \geq 0$, then $m \mapsto F(m,\e,T,m_0)$ is convex.
\bi
\item[(i)]
Note first that
 $
F(0,\e,T,m_0) = \frac{m_0}{(\e \,T +1)^2} > 0
$ 
(recall that here $m_0 \geq \e >0$). 
Moreover,
\[
F(\e,\e,T,m_0) = \frac{2(\e \,T + 1) \e \,T + m_0 \left[(\e \,T + 1) + \e^2 T^2\right]}{\left((\e\, T + 1)^2 - \e^2 T^2\right)^2} - \e = \psi(\e \,T,m_0) - \e,
\]
with
\[
\psi(y) := \frac{2(y+1)y + m_0 \left[(y+1)^2 + y^2 \right]}{(2y+1)^2}.
\]
As 
$
\psi'(y) = \frac{2(1-m_0)}{(2y+1)^3},
$ 
we deduce that $F(\e,\e,T,m_0)$ is strictly increasing in $T$, except for $m_0=1$ (in this case it is constant). In all cases we have that, since $F(\e,\e,0,m_0) = m_0-\e \geq 0$,
 $
F(\e,\e,T,m_0) >0,
$ 
$\forall T>0$. Thus, the map $m \mapsto F(m,\e,T,m_0)$ is strictly positive at the endpoints of the interval $(0,\e)$. We also have
 $
F(m,\e,0,m_0) = m_0 - m >0
$ 
$\forall m \in (0,\e)$. Moreover, it is easily seen that for each $m \in (0,\e)$,
\be{liminfT}
\lim_{T \ra +\infty} F(m,\e,T,m_0) = -m <0.
\ee
Then there must be a time $T_c^{(2)}(\e,m_0)$ such that for $T < T_c^{(2)}(\e,m_0)$, $F(m,\e,T,m_0) > 0$, $\forall m \in (0,\e)$, and the graph of the convex function $y = F(m,\e,T_c^{(2)}(\e,m_0),m_0)$ is tangent to the line $y=0$. Set
\[
\mathcal{T} := \left\{ T > 0 : \min_{m \in (0,\e)}F(m,\e,T,m_0) \leq 0 \right\}.
\]
The proof of this point (i) is completed as we show that $\mathcal{T} = [T_c^{(2)}(\e,m_0), +\infty)$. If this is not the case, by continuity, there must be a time $\hat{T} \geq T_c^{(2)}(\e,m_0)$,  $\hat{m} \in (0,\e)$ and $\d>0$ such that
\be{crit}
F(\hat{m},\e,\hat{T},m_0) = 0, \ \ \frac{d}{dm}F(\hat{m},\e,\hat{T},m_0) = 0,
\ee
but 
 $
F(\hat{m},\e,t,m_0)>0,
$ 
$\forall t \in (\hat{T}, \hat{T}+\d)$. To show that this is impossible, it is enough to prove that
\be{derneg}
\frac{d}{dT}F(\hat{m},\e,\hat{T},m_0) <0.
\ee
To see this it is convenient to perform the following change of variables: $u := \frac{m}{\e} \in (0,1)$ and $r := \e\, T$, so that
\be{changeofv}
F(m,\e,T,m_0) = G(u,\e,r,m_0) := \frac{1}{(1+r)^2} \varphi\left( \frac{r}{1+r} u, m_0\right) - \e u,
\ee
where $\varphi$ is given in \eqref{phifun}. Note that \eqref{derneg} is equivalent to
\be{derneg1}
\frac{d}{dr} G(\hat{u},\e,\hat{r},m_0) <0,
\ee
where $\hat{u} := \frac{\hat{m}}{\e}$ and $\hat{r} := \e \hat{T}$.  

\noindent
{\bf Claim}: $\hat{r} > \frac12$. 

{\it Proof.} To see this, note that, being $\e \leq m_0$,
\[
G(u,\e,r,m_0) \geq H(u,r,m_0) :=  \frac{1}{(1+r)^2} \varphi\left( \frac{r}{1+r} u, m_0\right) - m_0 u.
\]
The claim follows if we show that, $\forall r \leq  \frac12$,
\be{posr}
H(u,r,m_0) > 0, \ \ \forall u \in (0,1).
\ee
Since $H(u,r,m_0)$ is linear in $m_0$, it is enough to prove \eqref{posr} for $m_0\in\{0,1\}$. For $m_0 = 0$ this is obvious, so we show it for $m_0 = 1$:
\[
H(u,r,1) = \frac{\left(1+ \frac{r}{1+r} u \right)^2}{(1+r)^2 \left( 1- \frac{r^2 u^2}{(1+r)^2} \right)^2} - u = \frac{(1-u)(r^2 u^2 - (r^2 +2r)u+1)}{(1+r(1-u))^2},
\]
which has the same sign as 
 $
p(u) := r^2 u^2 - (r^2 +2r)u+1).
$ 
If $r \leq \frac12$ (indeed here $r<2$ would suffice), $p'(u) = 2 r^2 u - r^2 - 2r < 0$, $\forall u \in (0,1)$, so, 
$
p(u) \geq p(1) = 1-2r \geq 0,
$
$\forall u \in (0,1)$ and for $r \leq \frac12$. This completes the proof of the Claim.\qed

We are now left with the proof of \eqref{derneg1}. Note that
\be{derG}
\frac{d}{dr}G(u,\e,r,m_0) = - \frac{2}{(1+r)^3} \varphi\left( \frac{r}{1+r} u, m_0\right) + \frac{u}{(1+r)^4} \varphi'\left( \frac{r}{1+r} u, m_0\right),
\ee
where
$
 \varphi'(z,m_0) = \frac{d}{dz} \varphi(z,m_0).
$ 
 By \eqref{crit},
 \[
  G(\hat{u},\e,\hat{r},m_0) = 0 \ \Rightarrow \ \varphi\left( \frac{\hat{r}}{1+\hat{r}} \hat{u}, m_0\right) = \e \hat{u},
  \]
  and
\[
\frac{d}{du}G(\hat{u},\e,\hat{r},m_0) = 0 \ \Rightarrow \ \frac{r}{(1+r)^3} \varphi'\left( \frac{\hat{r}}{1+\hat{r}} \hat{u}, m_0\right) = \e.
\]
Inserting these identities in \eqref{derG} we obtain
\be{derG}
\frac{d}{dr}G(\hat{u},\e,\hat{r},m_0) = \frac{\e \hat{u}}{1+\hat{r}}\left(-2+\frac{1}{\hat{r}} \right) <0, \quad\text{ as } \hat{r} > \frac12.
\ee

\item[(ii)]
Note that $F(0,T,\e,m_0) = \frac{m_0}{(\e\, T+1)^2} >0$. Moreover,
 $
\lim_{T \ra +\infty} F(\e,T,\e,m_0) = \frac{1+m_0}{2} - \e \leq 0.
$ 
Since, as shown in point (i), the map $T \mapsto F(\e,T,\e,m_0) $ is strictly increasing, then
 $
 F(\e,T,\e,m_0) <0,
 $ 
 $\forall T >0$. Thus, $m \mapsto F(m,T,\e,m_0)$ has opposite sign at the endpoints of $(0,\e)$: by convexity there is a unique solution 
 \mbox{$m = M(T, \e,m_0)$}  to $F(m,T,\e,m_0) = 0$. The fact that 
 $
 \lim_{T \ra +\infty} M(T,\e,m_0) = 0
$ 
follows from the fact that 
 $
\lim_{T \ra +\infty} F(0,T,\e,m_0) = 0
$ 
and that 
$
\lim_{T \ra +\infty} \frac{d}{dm}F(0,T,\e,m_0)  = -1.
$ 
\item[(iii)]
In this case $F(0,T, \e,m_0) = \frac{m_0}{(\e\, T +1)^2} \geq 0$. Moreover, 
 $F(\e,T,\e,m_0) < 0 \ \iff \ T < T^*(\e,m_0),$
 as already seen in the proof of Proposition \ref{prop:equilibria}, point (iii). By convexity of $m \mapsto F(m,T,\e,m_0)$, for $T \leq T^*(\e,m_0)$ Equation \eqref{eq2alt} has a unique solution in $[0,\e)$. The fact that $m=0$ is a solution if and only if $m_0=0$ is easily verified. Moreover, by \eqref{liminfT}, for $T$ sufficiently large $F(m,T,\e,m_0)$ attains negative values, it is positive at $m=0$ and $m=\e$, so \eqref{eq2alt} has two solutions in $[0,\e)$. We need, however, a sharper analysis for $T> T^*(\e,m_0)$. Note that, in this case $F(0,T,\e,m_0) > 0$ and $F(\e,T,\e,m_0) >0$. Thus, by convexity of $m \mapsto F(m,T,\e,m_0)$,   \eqref{eq2alt} has zero or two solutions, except for the ``special times'' $T$ for which the graph of the map \mbox{$m \mapsto F(m,T,\e,m_0)$} is tangent to the horizontal axis. Note that these special times are identified as ($T$-component of the) solutions in $(0,\e) \times  (T^*(\e,m_0),+\infty)$ of the system

\be{syst}
\left\{
\begin{split}
F(m,T,\e,m_0) & =  0 \\
\frac{d}{dm}F(m,T,\e,m_0) & =  0.
\end{split}
\right.
\ee
\\
Note that \eqref{syst} has no solutions with $T \leq T^*(\e,m_0)$, so we may look for solutions of \eqref{syst} in $(0,\e) \times  (0,+\infty)$. The remaining part of the proof is based on the following Lemma.

\begin{lemma}\label{lemma:tec}
{\it Let $\e_*^{(2)}(m_0)$ and $\e_*^{(3)}(m_0)$ be as in the statement of Proposition \ref{prop:equilibria3}. Then,
\[
m_0 < \e_*^{(2)}(m_0) < \e_*^{(3)}(m_0) < \frac{1+m_0}{2},
\]
(unless for $m_0=0$, where $0 = \e_*^{(2)}(0) < \e_*^{(3)}(0) < \frac12$) such that
\bi
\item[(a)] 
for $m_0 < \e \leq \e_*^{(2)}(m_0)$, \eqref{syst} has a unique solution $(\tilde{m}(\e,m _0), \tilde{T}(\e,m_0))$;
\item[(b)]
for $\e_*^{(2)}(m_0) < \e <\e_*^{(3)}(m_0)$, \eqref{syst} has two solutions $(\hat{m}(\e,m _0), T_c^{(2)}(\e,m_0))$ and \\\mbox{$(\tilde{m}(\e,m _0), T_c^{(3)}(\e,m_0))$} with $ T_c^{(2)}(\e,m_0)<  T_c^{(3)}(\e,m_0)$;
\item[(c)]
for $ \e_*^{(3)}(m_0) \leq \e < \frac{1+m_0}{2}$, \eqref{syst} has no solutions.
\ei
}
\end{lemma}

The proof of this lemma is postponed after the end of this section. The desired result of the solutions of \eqref{eq2alt} readily follows from this Lemma. Indeed, in case (a), there is a unique special time $T_c^{(2)}(\e,m_0)$. Since, for large $T$, \eqref{eq2alt} has two solutions, necessarily for $T^*(\e,m_0) < T \leq T_c^{(2)}(\e,m_0)$ we must have $F(m,T,\e,m_0)>0$, \mbox{$\forall m \in (0,\e)$,} so \eqref{eq2alt}  has no solution. \\In case (b) there are two special times $ T_c^{(2)}(\e,m_0)<  T_c^{(3)}(\e,m_0)$: the only possibility is that, for $T^*(\e,m_0) < T < T_c^{(2)}(\e,m_0)$, the graph of $m \mapsto F(m,T,\e,m_0)$ crosses twice the horizontal axis (two solutions for \eqref{eq2alt}), for $T_c^{(2)}(\e,m_0)<
T<T_c^{(3)}(\e,m_0)$ it stays above the horizontal axis (no solutions for \eqref{eq2alt}) and it crosses again twice the horizontal axis for $T > T_c^{(3)}(\e,m_0)$. The proof is therefore completed.\qed
\ei

\subsection*{Proof of Lemma \ref{lemma:tec}}

By the change of variables $u := \frac{m}{\e} \in (0,1)$ and $r := \e\, T$ as in \eqref{changeofv}, we may replace $F(m,\e,T,m_0)$ by
\[
G(u,\e,r,m_0) := \frac{1}{(1+r+ru)^2} \left[ m_0 + \frac{2(1+m_0)(1+r)ru}{(1+r-ru)^2} \right] - \e u =0,
\]
$u \in (0,1)$, $r >0$, and \eqref{syst} is equivalent to
\be{syst1}
\left\{
\begin{split}
G(u,\e,r,m_0) & = 0 \\
\frac{d}{du} G(u,\e,r,m_0) & = 0.
\end{split}
\right.
\ee
Letting, as above, 
 $
U(u,\e,r,m_0):= m_0(1+r-ru)^2 + 2(1+m_0)r(1+r)u - \e u\left[(1+r)^2-r^2 u^2\right]^2 = 0,
$ 
we have that 
 $
G(u,\e,r,m_0) = \frac{1}{\left[(1+r)^2-r^2 u^2\right]^2} U(u,\e,r,m_0).
$ 
It is immediately seen that \eqref{syst1} is equivalent to
\be{syst2}
\left\{
\begin{split}
U(u,\e,r,m_0) & = 0 \\
\frac{d}{du} U(u,\e,r,m_0) & = 0,
\end{split}
\right.
\ee
We use again the identity 
 $
U(u,\e,r,m_0) = u \frac{d}{du} U(u,\e,r,m_0) +\left(m_0 -4\e r^2 u^3\right) \left((1+r)^2 - u^2 r^2 \right), 
$ 
which implies that \eqref{syst2} is equivalent to
\be{syst3}
\left\{
\begin{split}
\left(m_0 -4\e r^2 u^3\right) & = 0 \\
\frac{d}{du} U(u,\e,r,m_0) & = 0.
\end{split}
\right.
\ee
Summing up, the number of pairs $(m,T) \in (0,\e) \times (0,+\infty)$ solving \eqref{syst} equals the number of solutions $r$ to the equation
\be{syst4}
W(r,\e,m_0):= \frac{d}{du} U\left(\left(\frac{m_0}{4 \e r^2} \right)^{1/3},\e,r,m_0\right)  = 0
\ee
such that $\frac{m_0}{4 \e r^2} <1$, i.e., $r> \sqrt{\frac{m_0}{4\e}}$. With the help of a symbolic calculator we obtain
{\small{
\[
8W(r,\e,m_0) = r \left\{ 16 + \left[ 16 + 3m_0 \left(\frac{2m_0}{\e r^2} \right)^{1/3} \right] r \right\} + 4 \e (1+r)^2 \left\{ -2-4r+ \left[ -2 + 3 \left(\frac{2m_0}{\e r^2} \right)^{1/3}\right] r^2 \right\}.
\]
}}
Now set 
 $
V(s,\e,m_0) := W(\sqrt{s},\e,m_0).
$ 
Thus we are left with the problem of finding the solutions of 
\be{syst5}
V(s,\e,m_0) = 0
\ee
with $s > \frac{m_0}{4\e}$. This last change of variable is a trick to get the following claim.

\noindent
{\bf Claim} The map $s \mapsto V(s,\e,m_0)$ is strictly concave in $\left(\frac{m_0}{4\e},+\infty\right)$.

{\it Proof.} Using again a symbolic calculator we get, letting $k := \left(\frac{2m_0}{\e}\right)^{1/3}$,
\[
-12 s^{3/2} \frac{d^2}{ds^2}V(s,\e,m_0) = 6+m_0 k s^{1/6} + \e \left[ - 12 + 4 k^2 s^{-1/6} + 36 s + 5 k^2 s^{1/3} + 24 s^{3/2} - 8 k^2 s^{5/6} \right].
\]
To prove the claim we need to show that this last expression is positive. Note that $s > \frac{m_0}{4\e} = \left(\frac{k}{2}\right)^3$. 
Moreover, as $m_0 < \e < \frac{1+m_0}{2}$, then
 $
\frac{4m_0}{1+m_0} < k^3 < 2.
$ 
Again from $\e < \frac{1+m_0}{2}$, the inequality $ \frac{d^2}{ds^2}V(s,\e,m_0) <0$ follows if we show that
\be{d21}
6+m_0 k s^{1/6} + \frac{1+m_0}{2} \left[ - 12 + 4 k^2 s^{-1/6} + 36 s + 5 k^2 s^{1/3} + 24 s^{3/2} - 8 k^2 s^{5/6} \right] >0.
\ee
Using $s > \left(\frac{k}{2}\right)^3$ and $k^3 > \frac{4m_0}{1+m_0}$ we have that
 $
\frac{1+m_0}{2}5 k^2 s^{1/3} >\frac{1+m_0}{4}5k^3> 5m_0,
$ 
so the l.h.s. of \eqref{d21} is bounded from below by
\be{d22}
m_0 k s^{1/6} - m_0  + (1+m_0) \left[ 2k^2 s^{-1/6} + 18 s + 12 s^{3/2} - 4 k^2 s^{5/6} \right].
\ee
Now, for $s \leq \frac14$ the expression in \eqref{d22} is bounded from below by
\[
\begin{split}
- m_0 +  (1+m_0)  \left[ 2k^2 s^{-1/6} - 4 k^2 s^{5/6} \right] & = -m_0 +\frac{1+m_0}{s^{1/6}} \left(2k^2 - 4 k^2 s \right) \\ & \geq - m_0 +(1+m_0) k^2 s^{-1/6} \\ &  \geq  - m_0 +(1+m_0) \left(\frac{4 m_0}{1+m_0} \right)^{2/3} \\ & \geq - m_0 +  (1+m_0) 4^{2/3} \frac{m_0}{1+m_0} >0.
\end{split}
\]
For $s > \frac14$, the expression in \eqref{d22} is bounded from below by
\[
\begin{split}
- m_0 + 18 s + 12 s^{3/2} - 4 k^2 s^{5/6} & \geq -m_0 + s^{5/6} \left[ 18 \left(\frac14\right)^{1/6} + 12  \left(\frac14\right)^{2/3} - 4 k^2 \right] \\ & \geq  -m_0 + s^{5/6} \left[ \left(\frac14\right)^{1/6} + 12  \left(\frac14\right)^{2/3} - 4 \, 2^{2/3} \right] \\ & \geq -m_0 + 8s^{5/6} \geq -m_0 + 2 > 0.
\end{split}
\]
This proves \eqref{d21} and therefore the claim. \qed
\\

Now, using concavity of $V(s,\e,m_0)$ and the fact that 
 $$
\lim_{s \ra +\infty} V(s,\e,m_0) = \lim_{r \ra +\infty} W(r,\e,m_0) = -\infty,
$$
we have that \eqref{syst5} has a unique solution whenever 
 $
V\left(\frac{m_0}{4\e}, \e, m_0\right) >0.
$ 
We get
\[
V\left(\frac{m_0}{4\e}, \e, m_0\right) = \frac{m_0(1+m_0)}{2\e} + (1+m_0) \sqrt{\frac{m_0}{\e}} - \e\left(1+e \sqrt{\frac{m_0}{\e}} \right),
\]
so
\[
V\left(\frac{m_0}{4\e}, \e, m_0\right) \Big|_{\e = m_0} = \frac32(1-m_0) >0,
\]
\be{last}
V\left(\frac{m_0}{4\e}, \e, m_0\right) \Big|_{\e = \frac{(1+m_0)}{2}}  = -\frac12(1-m_0) < 0
\ee
and
\[
\frac{d}{d\e} V\left(\frac{m_0}{4\e}, \e, m_0\right) = -1- \sqrt{\frac{m_0}{\e}}-\frac{m_0(1+m_0)}{2\e^2} - 
\frac{ \sqrt{\frac{m_0}{\e}}(1+m_0)}{2\e} < 0.
\]
Therefore, there is a unique $\e_*^{(2)}(m_0)$, with $m_0 \!<\! \e_*^{(2)}(m_0) \!<\! \frac{1+m_0}{2}$, such that \mbox{$V\!\left(\frac{m_0}{4\e_*^{(2)}}, \e_*^{(2)}, m_0\right) \!\!=\! 0$.} Moreover, for $m_0 < \e < \e_*^{(2)}(m_0)$, \eqref{syst5} has a unique solution and, since
\[
\frac{d}{ds} V(s,\e,m_0) \Big|_{s = \frac{m_0}{4\e}} = \frac{2(1+m_0-2\e) \left(1+\sqrt{\frac{m_0}{\a}}\right)}{\sqrt{\frac{m_0}{\a}}} >0,
\]
it follows that \eqref{syst5} has two solutions as $\e$ crosses $\e_*^{(2)}(m_0)$. This actually occurs until $\e$ reaches $\e_*^{(3)}(m_0)$, where  $\e_*^{(3)}(m_0)$ is characterized by the fact that the graph of $s \mapsto V(s,\e_*^{(3)}(m_0),m_0)$ is tangent to the horizontal axis. This is a consequence of the following monotonicity property: 
\be{mon}
\frac{d}{d\e} V(s,\e,m_0) <0,\quad \forall s \geq \frac{m_0}{4\e}.
\ee
This suffices to characterize $\e_*^{(3)}\!(m_0)$; to complete the proof we need to show that \mbox{$\e_*^{(3)}\!(m_0) \!<\! \frac{1+m_0}{2}$} which is equivalent to
\be{mon1}
V\left(s,\frac{1+m_0}{2},m_0\right) <0,\quad\forall s \geq \frac{m_0}{2(1+m_0)}.
\ee
We are therefore left to prove \eqref{mon} and \eqref{mon1}. We begin with \eqref{mon}.
\[
\frac{d}{d\e} V(s,\e,m_0) = -\frac{1}{8\e} \left[ m_0 \left(\frac{2m_0}{\e}\right)^{1/3} s^{2/3} + 4 \e(1+\sqrt{s})^2 \left( 2 + 4 \sqrt{s} + 4 s -\left(\frac{2m_0}{\e}\right)^{2/3} s^{1/3}\right)\right].
\]
To show that this expression is negative we just observe that, being $\frac{2m_0}{\e} \leq 2$,
\[
2 + 4 \sqrt{s} + 4 s -\left(\frac{2m_0}{\e}\right)^{2/3} s^{1/3} > 2 + 4 \sqrt{s} + 4 s -2 s^{1/3} > \left\{ \begin{array}{ll} 2-2s^{1/3} \geq 0 & \mbox{for } s \leq 1 \\ 4s - 2 s^{1/3} > 0 & \mbox{for } s>1. \end{array} \right.
\]
This establishes \eqref{mon}. Now we show \eqref{mon1}. We recall that $s \mapsto V\left(s,\frac{1+m_0}{2},m_0\right)$ is strictly concave for $s \geq \frac{m_0}{2(1+m_0)}$. Moreover, again with the help of a symbolic calculator
\[
\frac{d}{ds} V\left(s,\frac{1+m_0}{2},m_0\right) \Big|_{s = \frac{m_0}{2(1+m_0)}} = 0.
\]
So it is enough to show that
\[
V\left(\frac{m_0}{2(1+m_0)},\frac{1+m_0}{2},m_0\right) <0,
\]
that has been seen already in \eqref{last}. The proof is now complete.\qed

\subsection*{Proof of Proposition \ref{prop:equilibria4}: unpolarized incoherent equilibria ($-\e \leq m < 0$)}

As done in Proposition \ref{prop:equilibria2}, we use the symmetry \eqref{simmF}. So we look for solutions $m \in (0,\e]$ of the equation
 $
F(m,\e,T,-m_0) = 0;
$ 
this allows to reuse some of the ideas in Proposition \ref{prop:equilibria3}.
\bi
\item[(i)]
We employ here the change of variables seen in \eqref{changeofv}, so
\[
\begin{split}
F(m,\e,T,-m_0) & = G(u,\e,r,-m_0) := \frac{1}{(1+r)^2} \varphi\left( \frac{r}{1+r} u, -m_0\right) - \e u \\ & = \frac{2(1+r)ru-m_0\left[(1+r)^2+r^2 u^2\right]}{\left((1+r)^2-r^2 u^2\right)^2} - \e u \\
& = \frac{1}{(1+r+ru)^2} \left[ -m_0 + \frac{2(1-m_0)(1+r)ru}{(1+r-ru)^2} \right] - \e u.
\end{split}
\]
We need to show that this last expression is strictly negative $\forall u\in(0,1)$; it is enough to show this for $\e = \frac{1-m_0}{2}$. This amounts to prove that
\[
-m_0 +  \frac{2(1-m_0)(1+r)ru}{(1+r-ru)^2} < \frac{1-m_0}{2} u (1+r+ru)^2,
\]
which follows from
\[
\frac{2(1-m_0)(1+r)ru}{(1+r-ru)^2} < \frac{1-m_0}{2} u (1+r+ru)^2 \ \iff \ 4r(1+r) < \left[(1+r)^2 - r^2 u^2 \right]^2,
\] 
$\forall u \in (0,1)$. This last inequality follows if we show it holds for $u=1$, i.e.,
 $ 
4r(1+r) < (1+2r)^2,
$ 
that is clearly true $\forall r>0$.

\item[(ii)]
As seen in the proof of point (i) of Proposition \ref{prop:equilibria3}, $F(\e,\e,T,-m_0)$ is strictly increasing in $T$. Moreover, the same computation done after \eqref{eq0} shows that $F(\e,\e,T,-m_0)>0$ if and only if $T>T^*(\e,-m_0)$. By the same change of variables used in point (i), \eqref{eq2alt} is equivalent to the equation
\[
\frac{1}{(1+r+ru)^2} \left[ -m_0 + \frac{2(1-m_0)(1+r)ru}{(1+r-ru)^2} \right] - \e u =0,
\]
with $u \in (0,1)$, which is also equivalent to
\[
U(u,\e,r,m_0):= -m_0(1+r-ru)^2 + 2(1-m_0)r(1+r)u - \e u\left[(1+r)^2-r^2 u^2\right]^2 = 0.
\]
Our proof is based on the following claim.

\noindent
{\bf Claim}. Let $u^* \in (0,1)$ be such that $U(u^*,\e,r,m_0) = 0$. Then, 
 $
\frac{d}{du} U(u^*,\e,r,m_0)>0.
$ 
\\

{\it Proof.} We use the identity 
\[
U(u,\e,r,m_0) = u \frac{d}{du} U(u,\e,r,m_0) +\left(-m_0 -4\e r^2 u^3\right) \left((1+r)^2 - u^2 r^2 \right) < u \frac{d}{du} U(u,\e,r,m_0)
\]
which proves the claim. \qed 

This clearly implies that such $u^*$ exists if and only if $U(1,\e,r,m_0) >0$ (as $U(0,\e,r,m_0) = -m_0(1+r)^2\leq 0$), i.e., if and only if $r > \e\, T^*(\e,-m_0)$ and in this case it is unique. This completes the proof.\qed
\ei

\bibliographystyle{}

\end{document}